\documentclass[a4,10pt]{article}

\usepackage{graphicx}
\usepackage{natbib}
\usepackage{psfrag}
\usepackage{bm}
\usepackage{color}
\usepackage{amsmath}
\usepackage{amssymb}
\usepackage{amsfonts}
\usepackage{natbib}
\usepackage{psfrag}
\usepackage{fancyhdr}
\usepackage{float}

\newsavebox{\astrutbox}
\sbox{\astrutbox}{\rule[-5pt]{0pt}{20pt}}

\begin{document}
\fancyhf{}
\pagestyle{fancy}
\rhead{\textit{Why, How and when MHD turbulence becomes 3D}}
\lhead{\textit{A. Poth\'erat \& R. Klein}}

\title{Why, how and when MHD turbulence at low $Rm$ becomes \emph{three}-dimensional}

\author{A. Poth\'erat$^1$ and R. Klein$^2$\\
$^1$Applied Mathematics Research Centre, \\
Coventry University, Coventry CV5 8HY, UK\\
$^2$Technische Universit\"at Ilmenau, Fakult\"at f\"ur Maschinenbau,\\
 Postfach 100565, 98684 Ilmenau Germany
}%


\date{21$^{st}$ October 2014}
\maketitle
\cfoot{\thepage}
\renewcommand{\headrulewidth}{0pt}

\begin{abstract}
{Magnetohydrodynamic (MHD) turbulence at low Magnetic Reynolds
number is experimentally investigated by studying a liquid metal flow
in a cubic domain. We focus on the mechanisms that
determine whether the flow is quasi-two dimensional, three-dimensional or in any intermediate state. To this end,
forcing is applied by injecting a DC current $I$ through one wall of the cube only,
to drive vortices
spinning along the magnetic field. Depending on the intensity of the externally applied magnetic field, these vortices extend part or all of the way through the cube.
Driving the flow in this way allows us to precisely control not only the forcing intensity
but also its dimensionality.
A comparison with the theoretical analysis of this configuration singles out
the influences of the walls and of the forcing on the flow dimensionality}. Flow dimensionality is characterised in several ways. First, we show that when
inertia drives three-dimensionality, the velocity near the wall where current
is injected scales as $U_b\sim I^{2/3}$. Second, we show that when the
distance $l_z$ over which momentum diffuses under the action of the Lorentz
force (\cite{sm82}) reaches the channel width $h$, the velocity near the opposite wall $U_t$ follows a similar law with a correction factor $(1-h/l_z)$ that measures
three-dimensionality. When $l_z<h$, by contrast, the opposite wall has less
influence on the flow and $U_t\sim I^{1/2}$. The central role played by the ratio
$l_z/h$ is confirmed by experimentally verifying the scaling $l_z\sim N^{1/2}$
put forward by \cite{sm82} ($N$ is the interaction parameter) and finally,
the nature of the three-dimensionality involved is further clarified
by distinguishing \emph{weak} and \emph{strong}
three-dimensionalities previously introduced by \cite{kp10_prl}. It is found
that both types vanish only asymptotically in the limit $N\rightarrow\infty$.
This provides evidence that because of the no-slip walls, 1) the transition between
quasi-two-dimensional and three-dimensional turbulence does not result from a
global instability of the flow, unlike in domains with non-dissipative boundaries
(\cite{boeck08_prl}), and 2) it doesn't occur simultaneously at all scales.
\end{abstract}


\section{Introduction}
Magnetohydrodynamic (MHD) turbulence at low magnetic Reynolds number has a well-known  tendency to become two-dimensional. In this paper, we experimentally characterise the mechanisms through which two-dimensionality breaks down in these flows and quantify the ensuing three-dimensionality. 
Besides the fundamental question of understanding this type of turbulence, 
liquid metal flows, in which MHD turbulence is usually found, are of central industrial interest. In  the metallurgical and nuclear sectors, they are either processed or used to carry heat and mass. 
Typical examples include 4$^{\rm th}$ generation sodium nuclear fission reactors and the cooling blankets of nuclear fusion 
reactors  (\cite{vecha13_pf}). Since two- and three-dimensional turbulence have radically 
different transport and dissipation properties, the question of the 
dimensionality of turbulence is key to the efficiency of these systems. In such 
engineering configurations, the flow is neither intense nor electrically 
conductive enough to advect the magnetic field over a short enough time to compete with 
magnetic diffusion. This justifies their description in the frame of the 
low-$R\!m$ approximation, where the coupling between electrical and mechanical 
quantities reduces to that between velocity, pressure and local electric 
current density (\cite{roberts67}). Under this 
approximation, \cite{sm82} showed in a seminal paper to which the title of this work pays homage, that the main effect of the Lorentz force was to diffuse the 
momentum of a structure of size $l_\perp$ along the externally applied magnetic 
field $\mathbf B$ in time $\tau_{2D}=(\rho/\sigma B^2)(l_z/l_\perp)^2$ over a distance $l_z$ 
($\rho$ and $\sigma$ are the fluid density and electric conductivity). This effect explains the tendency of these flows to two-dimensionality when the field is 
homogeneous. When turbulence is present, diffusion is achieved over a distance 
$l_z^{(N)}\sim N^{1/2}$, where the interaction parameter $N$ is the ratio of 
the eddy turnover time $\tau_U(l_\perp)=l_\perp/U(l_\perp)$ 
to the Joule dissipation time $\tau_J=\rho/\sigma B^2$. This remarkable 
property was exploited 
extensively to reproduce some of the fine properties of two-dimensional 
turbulence, such as the inverse energy cascade, in thin {horizontal} 
layers of liquid metal pervaded by a strong {vertical} magnetic field (\cite{sommeria86, sommeria88}, see figure \ref{fig:3dsketch} for a generic representation of this geometry). 
It also places MHD flows in the much wider class of flows with a tendency to two-dimensionality: this class includes flows in a background rotation (\cite{greenspan69}) or where 
stratification is present (\cite{paret97_pf}). Geophysical flows (oceans and atmospheres) are famous examples. The question of the dimensionality of turbulence in plane fluid layers is key to understanding the dynamics of all such systems.\\ 
In low-$R\!m$ MHD, the question of dimensionality emerged with numerical simulations in bounded domains: 
\cite{schumann76} showed the first evidence of MHD turbulence becoming 
strictly two-dimensional in a three-dimensional periodic domain when 
$N$ was significantly greater than unity, 
and this was later confirmed by \cite{zikanov98_jfm}.
More recently, \cite{boeck08_prl} found, still numerically, that at moderate 
values of $N$, the flow could shift
intermittently between two- and three-dimensional states.
\cite{thess07_jfm} found the same phenomenon in ellipsoidal {structures  confined} by slip-free boundaries,
while \cite{pdy10_jfm} showed that for a flow in a given magnetic field,
three-dimensionality appeared at a bifurcation when the intensity of a {two-dimensional flow with a velocity field orthogonal to the magnetic field (2D-2C 
flow)}, was increased. All these studies with dissipation-free
boundaries drew a picture where three-dimensionality developed as an
instability on initially two-dimensional (and two-component) flows.\\
By contrast, strict two-dimensionality cannot be achieved when physical walls 
are present, as in experiments or in oceans and planetary atmospheres, because 
of the three-dimensionality of wall boundary layers. Furthermore, friction there 
drives mass or electric current (respectively in rotating 
and MHD flows), that recirculate in the core under the form of Ekman pumping or 
eddy currents. Owing to this effect, transversal components of velocity or electric current almost always subsist in thin layers of fluid, even when weak boundary 
friction is present. This was observed both in non-MHD (\cite{akkermans08_epl}), and MHD flows. Not only does it affect the quasi-two-dimensional dynamics of the flow but it also induces three-dimensionality in the 
core (\cite{albouss99,psm00_jfm, psm05_jfm}). Asymptotic analyses indeed 
predict a quadratic variation of velocity across the bulk of the fluid layer, 
called the \emph{Barrel effect} (\cite{psm00_jfm,p12_epl}), which was observed in 
the numerical simulations of \cite{muck00_jfm}.\\
{An inhomogeneous forcing across the fluid layer was shown to drive
three-dimensionality too, and the resulting three-dimensional
structures were recently found to induce secondary flows that could override Ekman 
pumping (\cite{prcd13_epje}). These results concur to show that whether induced  by the boundaries or by the inhomogeneity of the forcing, 1) the appearance of three-dimensionality {and the appearance of the third velocity component} are interdependent, and 2) three-dimensionality does not always result from instabilities, at least in wall-bounded flows. Our previous experiments in fact suggest that 
three-dimensionality due to recirculating flows or electric currents 
and instability-driven three-dimensionality could co-exist, but tended to take distinct forms: 
respectively a \emph{weak} form, where only the flow intensity varies along $\mathbf B$ and 
a \emph{strong} {form} where flow topology varies too (\cite{kp10_prl}). 
Recently, it was also 
shown that because three-dimensional flows are more dissipative, they carry less energy than their quasi-two dimensional counterpart for the same level of external forcing and they also decay faster when not forced. This property was used to detect three-dimensionality in non-MHD flows (\cite{shats10_prl, duran-matute10_pre}).\\
}
Our purpose is to characterise three-dimensionality in low-$R\!m$ MHD 
turbulence in wall-bounded fluid layers and to determine its driving mechanism. Our starting point shall be to quantify how flow intensity varies with dimensionality, somewhat in the spirit of \cite{duran-matute10_pre}. 
Our approach consists of forcing turbulence 
between two walls orthogonal to an externally imposed magnetic field 
in a thick enough layer of fluid  to observe how three-dimensionality develops.
We shall first describe how the experimental set-up can reproduce the generic properties of MHD turbulence in channels (section \ref{sec:setup}). We shall then establish theoretical scalings linking near-wall velocities to 
the intensity of the electric current driving flows in MHD channels
(section \ref{sec:theory}). Occurrences of three-dimensionality will then be 
tracked in the light of these scalings (section \ref{sec:visc-inertia}) and 
an experimental measure of the corresponding values of $l_z^{(N)}$ shall be obtained (section \ref{sec:wall_scalings}). 
This picture shall finally be refined by means of frequency analysis and through a quantitative measure of weak and strong three-dimensionalities (section \ref{sec:strong_weak}).
\section{\label{sec:setup}The \emph{FLOWCUBE} Experimental facility}
\subsection{\label{sec:m_desc}Mechanical description}
\begin{figure}
\centering
\includegraphics[width=10.5cm]{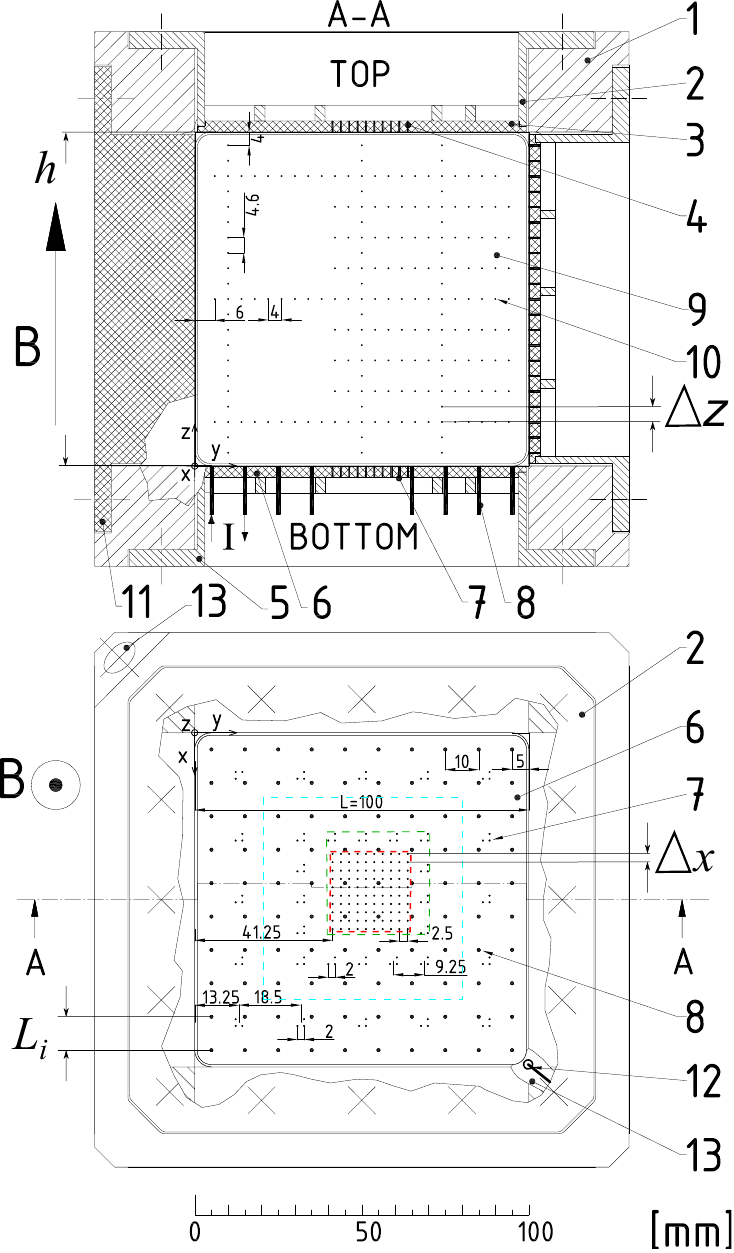}
\caption[Sketch of the experiment to study the appearance of three-dimensionality in wall-bounded magnetohydrodynamic flows.]{Sketch of the cubic container. Top: cross section. Bottom: top view onto the bottom plate. (1) cubic brazen frame; (2) top plate with (3) top electronic board (top Hartmann wall); (4) set of $196$ potential probes from the top Hartmann wall ; (5) bottom plate with (6) bottom electronic board (bottom Hartmann wall) ;(7) set of $196$ potential probes from the bottom Hartmann wall (as on the top plate) and (8) additional $100$ forcing electrodes ; (9) electronic board at the Shercliff wall with (10) $195$ potential probes 
; (11) side plate; (12) probe for electric potential reference; (13) inlet and outlet to evacuate the container and to fill it with liquid metal. All container walls are electrically insulating, except at the locations of potential probes and current injection electrodes. The central square region with high density of electric potential probes on the Hartmann walls is marked with a red dashed 
line.}
\label{fig:design}
\end{figure}

\begin{figure}
\centering
\includegraphics[width=14cm]{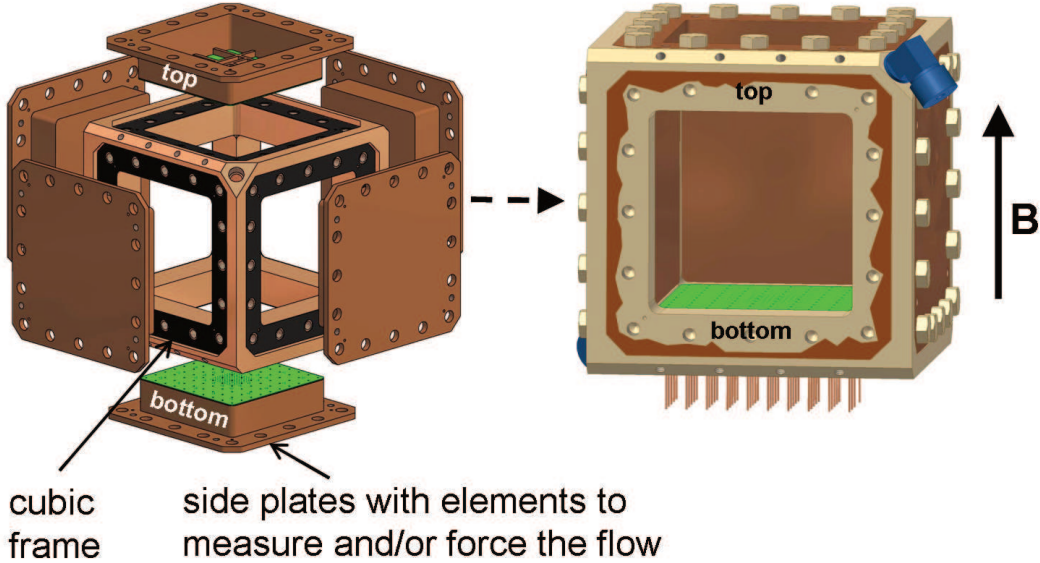}
\psfrag{E}{\begin{small} electrically \end{small}}
\psfrag{E1}{\begin{small} connected  \end{small}}
\psfrag{E2}{\begin{small} to the electrodes \end{small}}
\psfrag{C}{\begin{small} copper bars \end{small}}
\psfrag{P}{\begin{small} aluminium cooling plate \end{small}}
\psfrag{R}{\begin{small} $R_{2\Omega}$ resistance \end{small}}
\psfrag{U}{\begin{small} U-shaped connector \end{small}}
\psfrag{S}{\begin{small} switchboard \end{small}}
\psfrag{V}{\begin{small} DC power supply \end{small}}
\caption{ 
Modular design based on a cubic frame with interchangeable side plates. These 
plates can be easily swapped to exchange flow forcing and flow measurement 
systems. Left: open 
container with side plates unmounted. Right: closed container where the front 
plate is only partly represented to allow for a view into the container.
}
\label{fig:switch}
\end{figure}

The principle of the experiment follows that of \cite{sommeria86} 
on quasi-two dimensional turbulence in which a constant, almost homogeneous 
magnetic field $\mathbf{B}=B\mathbf e_z$ was applied across a square, shallow container of 
height $0.02$ m filled with liquid mercury (the frame origin is chosen at the 
centre of the bottom wall, with $\mathbf e_x$ and $\mathbf e_y$ pointing along its edges). References to top and bottom walls are used for convenience, since 
gravity plays no relevant role in this experiment. In {this} 
configuration, the 
time scale for two-dimensionalisation $\tau_{2D}(l_\perp)$ for each flow 
structure of transverse size $l_\perp$ was less than $10^{-2}$s  and much 
smaller than its turnover time $\tau_U(l_\perp)$. The flow was assumed quasi 
two-dimensional on these grounds. 
Unlike this earlier experiment though, 
the present container is not shallow, but cubic with inner edge $h=0.1$ m (Fig. \ref{fig:design}). Walls are impermeable and electrically insulating, except where small electrodes 
are inserted, either to drive or diagnose the flow. Since Hartmann and Shercliff 
layers are expected to develop along walls perpendicular and parallel to 
$\mathbf B$, we shall refer to them as \emph{Hartmann} walls and 
\emph{Shercliff} walls, respectively (Fig.~\ref{fig:design} and 
Fig.~\ref{fig:switch}). For the lower magnetic fields used, 
$\tau_{2D}(l_\perp)$ is of the order of $1s$. Such times are much longer than 
those of \cite{sommeria86}, and potentially comparable to, or longer than eddy 
turnover times, so three-dimensionality is expected to be present.
The working liquid metal is Gallinstan (or MCP11), an 
eutectic alloy of gallium, 
indium and tin that is liquid at room temperature, with electric conductivity 
$\sigma=3.4 \times 10^6$ S/m, density $\rho= 6400$ kg/m$^3$ and viscosity 
$\nu = 4 \times 10^{-7}$ m$^2$/s.
The container is successively filled with argon and {evacuated}
 several times
so as to ensure that as little gas as possible remains inside the vessel, and 
to prevent oxidation of the Gallinstan as much as possible. This way, the liquid 
metal 
is in the tightest mechanical contact with the walls and in electrical contact with all 
wall-embedded electrodes. Once filled with Gallinstan, the container is 
subjected to a magnetic field $B \mathbf e_z$ by being placed at the centre 
of the cylindrical bore of a superconducting solenoidal magnet.
Magnetic fields $B \in [0.1, 5]$T of a maximum inhomogeneity of 
$3\%$ along $\mathbf e_x$, $\mathbf e_y$,  and $\mathbf e_z$ and with 
corresponding Hartmann numbers $H\!a=Bh(\sigma/(\rho\nu))^{1/2}$ $\in [364, 18220]$ are achieved ($H\!a^2$ measures the ratio of Lorentz to viscous forces, see section \ref{sec:theory}). 
This very low level of inhomogeneity is not expected to have any significant 
effect on the measurements. By contrast, a possible curvature of the magnetic 
field lines due to the bipolar structure of the magnetic field could have an 
influence on the diagnosis of three-dimensionality. By comparing the position 
of several vortex centres in a steady quasi-two dimensional regime using interpolation between probes, we found that the lateral 
streamline displacement between top and bottom walls was approximately the size 
of an injection electrode (1 mm), which 
is below the spatial resolution of our measurements. The lack of influence of 
inhomogeneity was also confirmed by the very low level of three-dimensionality 
 we measured in the asymptotic quasi-two dimensional regime (section 
\ref{sec:strong_weak}).\\
%
The flow entrainment relies on the same principle as in 
\cite{sommeria86} and \cite{kpa09_pre}'s experiments. A DC electric current in the range 
$I \in [0-300]$A is injected at the bottom \textit{Hartmann} wall located at 
$z=0$ (Fig.~\ref{fig:design}), through a lattice of either $n=100$ or $n=16$ 
electrodes, each of diameter $1$ mm. The basic forcing geometry consists 
of a 
square array of either $10\times10$ or $4\times4$ electrodes 
respectively spaced by distances of either $L_i =0.01$ m or $L_i=0.03$ m
(non-dimensionally, $\lambda_i=L_i/h=0.1$ or  $\lambda_i=0.3$), both 
centred on the bottom Hartmann plate. For high enough injected current,
this basic pattern turns into turbulence, with adjustable forcing scale
$\lambda_i$. The electrodes are made of  copper and 
gold plated, which ensures provision of a good 
electrical contact with the Gallinstan. They are embedded into the bottom 
\textit{Hartmann} wall located at $z=0$, and mounted flush so that they do not 
protrude into the liquid metal and bring no mechanical disturbance into the 
flow. 
At their other end, they are connected to a regulated DC power supply through a 
dedicated switchboard system. 
This system makes it possible to connect 
each of the 100 available electrodes to either the positive or the negative 
pole of the DC power supply or to leave it open-ended. The 10x10 and 4x4 arrays 
are obtained by alternately connecting all the electrodes to either pole ($10\times10$), or by leaving 2 unconnected electrodes between one connected to the positive pole and one connected to the negative pole ($4\times4$). For low 
 injected electric current $I$, the base flow consists of vortices of 
diameter $L_i$ spinning in alternate directions as in \cite{sommeria86}'s 
experiment.
{This type of setup and forcing allowed \cite{sommeria86} to 
provide one of the first
experimental evidence of the inverse energy cascade, a distinctive feature of
two-dimensional turbulence. In that sense, the constant periodic forcing 
generates quasi-two dimensional turbulence at high $N$ when the injected current sufficiently exceeds the critical value for the destabilisation of the 
array. It should also be noted that a constant forcing is necessary to study 
the dimensionality of MHD turbulence as we set out to do. Without it, 
three-dimensionality would be suppressed by the action of the 
Lorentz force and turbulence would decay over timescales of the order of 
$\tau_{2D}(L_i)$ or of the Hartmann friction time, depending on whether it is 
 three- or quasi-two dimensional (\cite{sm82}).}\\ 
Since all electrodes are connected to a single power supply, particular 
attention was paid to ensure that the same, constant current intensity $I$ 
passed through all of them, despite random fluctuations in the contact 
resistance between the electrode surface and the liquid metal. 
These fluctuations of amplitude of the order 
of $10^{-2}$ $\Omega$ can lead to an irregular current distribution resulting 
into an ill-controlled forcing. By adding a constant ohmic resistance of 
$2\Omega\pm 0.25\%$ to all $100$ electric circuits, the impact of these fluctuations 
on the uniformity of the forcing becomes negligible. 
Each $2\Omega$ resistance dissipates a maximal power of $100$W, which limits the 
injected electric current per electrode to about $7$A. To evacuate heat, they 
are mounted on water-cooled aluminium plates. {The current injected in each electrode is monitored prior to each acquisition and discrepancies between the values measured at each electrode were found to remain on the order of $0.25\%$.}\\
%
%
%
%
\subsection{\label{sec:m_tech} Electric potential measurements}
The flow is analysed by measuring the electric potential $\phi$ locally, at probes 
that are embedded in the container walls. Each electric potential is 
obtained with respect to a reference taken in the thin inlet pipe located in a 
corner of the vessel, where the liquid metal is always at rest 
(see Fig.~\ref{fig:design}). Each probe consists of a gold plated copper wire 
 0.2 mm in diameter fitted flush to the container wall, in the same way as 
the current injection electrodes are. This way, {they incur practically 
no mechanical nor electrical disturbance on the flow.}
 196 of these probes are embedded in the $x-y$ plane, in the bottom $(z=0)$ and top $(z=h)$ \textit{Hartmann} walls, and 195 probes in the $y-z$ plane at the 
Shercliff wall at $x =0$. The sets of potential probes embedded in the top 
and bottom 
\textit{Hartmann} walls are mirror symmetric and precisely aligned  
opposite each other along magnetic field lines. Near the centre of the 
Hartmann walls, 
probes are positioned in a dense $10\times 10$ grid of spacing $\Delta x=\Delta y=\Delta=2.5$mm to map the smaller spatial variations of potential. We shall refer to this region of the $(x,y)$ plane as the "central square" (red dashed line in Fig. \ref{fig:design}). Mid-size 
and box-size structures are captured by measurements on sets of three probes 
(distant by $\Delta x=\Delta y =2$ mm from one another so as to record 
local variations of $\phi$ along $\mathbf e_x$ and $\mathbf e_y$), around the 
centre array and further out close to the Shercliff walls  
(Fig.~\ref{fig:design}). On the 
Shercliff wall at $x=0$, probes are respectively spaced by $\Delta y=4$mm and 
$\Delta z = 4.6$mm along $\mathbf e_y$ and $\mathbf e_z$. Signals recorded 
there give an insight on the structure of the flow along the magnetic field 
lines (see section \ref{sec:wall_scalings}). \\
{The signals are collected via a printed circuit leading to 
16-pin connectors built-in at the back of all walls. From there, the signal is 
conveyed to} a high-precision single-ended 736-channel acquisition system, 
manufactured and tailored to this particular experiment by company 
NEUROCONN ({http://www.neuroconn.de/profile/}). Since the signals can be 
of the order of $10\mu$V, each channel features a low-noise amplifier with 
gain  111. All signals are then synchronously sampled to \emph{$24$-bit} 
precision at a frequency of $128$Hz. Digital signals are optically 
transmitted to a PC where they are recorded through a dedicated 
MATLAB/SIMULINK module. Since frequencies relevant to the flow are typically 
expected in the range $[0,35]$ Hz, each channel is fitted with a low-pass filter 
of cut-off frequency of $45$ Hz. The peak-to-peak background noise that remains on the filtered signal is less than 2$\mu$V, and this determines the precision 
of the measurement system. {The high dynamic range dedicated analogue-to-digital  (A/D) conversion on each channel combined with the absence of detectable drift 
in the signals guarantees that no loss in precision occurs when subtracting 
the signals from neighbouring probes to construct gradients of electric 
potential}.\\ 
The measurement chain provides real-time simultaneous time series of electric 
potential in 587 locations spread between the top and bottom Hartmann walls and 
on one of the Shercliff walls (thereafter denoted $\phi_b$ and $\phi_t$ and 
$\phi_S$). At the Hartmann walls, gradients of electric potential at $\mathbf r$ shall be determined from a second order approximation:
\begin{equation}
\nabla_\perp\phi(\mathbf r)=\frac1\Delta\left[
\begin{array}{c}
{\phi(\mathbf r+\frac\Delta2\mathbf e_x)-\phi(\mathbf r-\frac\Delta2\mathbf e_x)}\\
{\phi(\mathbf r+\frac\Delta2\mathbf e_y)-\phi(\mathbf r- \frac\Delta2\mathbf e_y)}
\end{array}
\right]+\mathcal O(\Delta^3).
\label{eq:gphi_discrete}
\end{equation}
Electric potentials measured on Hartmann walls can also be related to the 
velocity fields $\mathbf u_b$ and $\mathbf u_t$ in the bulk of the flow just 
outside the Hartmann 
layers using Ohm's law there, and neglecting  variations of potential across the 
Hartmann layers, which are of the order of $Ha^{-2}$. For the bottom wall, 
\begin{equation}
\mathbf u_b={B^{-1}}\mathbf e_z\times\nabla_\perp\phi_b +(\sigma B)^{-1}\mathbf e_z\times\mathbf J_b,\\
\label{eq:u_exp_j}
\end{equation}
where $\mathbf J_b$ is the current density just outside the Hartmann layer.
In quasi-two-dimensional flows, $J_b\sim \sigma BUH\!a^{-1}$ can be neglected. 
(\ref{eq:u_exp_j}) then provides an indirect measurement of the velocity field, 
which has been widely used in MHD experiments featuring thin fluid layers (\cite{albouss99}).
\cite{kljukin98_ef} further noticed that in this case, the streamfunction in the bulk becomes proportional to $\phi_b$ and so isolines of $\phi_b$ provide 
a direct visualisation of flow patterns. When the flow becomes three-dimensional, however, $J_b$ becomes larger but still depends on the mechanism which pulls
current  in the core. For instance, we shall see in section \ref{sec:theory} that when 
inertial effects are present, $J_b=\mathcal O(N^{-1})$, so even in this case, flow 
patterns can still be identified and velocities estimated from (\ref{eq:u_exp_j}), albeit with a precision that decreases with $N^{-1}$ (see \cite{sreenivasan02}, where errors incurred by this measurement technique are estimated). We 
shall follow this strategy 
to test the scaling 
laws derived in section \ref{sec:theory} against our measurements. We shall 
also  obtain an alternative, more precise quantification of 
three-dimensionality, using potential measurements directly in section 
\ref{sec:strong_weak}. For this, we shall rely on the property that  
in a quasi-two dimensional flow, measurements of electric potentials and their 
gradient taken at top and bottom Hartmann 
walls should be fully correlated, whereas any trace of three-dimensionality 
should translate into a loss of correlation between these quantities. 

\subsection{Experimental procedure and strategy}
Experiments are performed at a constant magnetic field (constant $Ha$). Starting
 from a flow at rest, the current per electrode, (measured non-dimensionally by parameter $Re^0$ defined in section \ref{sec:theory}), is increased in about 20 
steps, spread logarithmically between 0 and 6A. 
At each step, 
electric potentials were recorded in a statistically steady state, which was 
deemed reached after typically $5\tau_H$, where the Hartmann friction time 
$\tau_H=h^2/(2\nu H\!a)$, which characterises quasi-dimensional dynamics 
(\cite{sm82}), is much longer than $\tau_J$ which characterises three-dimensional effects.
Records extended over a maximum of 7 to 10 
min, imposed by limitations of the cooling system for the superconducting magnet. 
To ensure stability of the magnetic field, the magnet was operated in 
"persistent mode", \emph{i.e.} with the coil disconnected from its power 
supply. Variations of magnetic field remained unmeasurable over a much longer 
time than the few consecutive hours needed to record data at one given field.\\
Forcing the flow at one wall only imposes a particular topology to the forcing 
that allows us to derive crucial generic information on the properties of
MHD turbulence in channels. First, the flow is forced near the bottom wall, where current is injected, but free near the top wall. Comparing flows in the
vicinity of both walls therefore gives us a reliable way to quantify three-dimensionality. Second, the influence of the top wall depends only on the dimensionality of turbulence and should therefore reflect the generic properties of walls in MHD turbulence in channels (section \ref{sec:visc-inertia}). Finally, by analysing electric
potentials along $\mathbf B$ at the side wall, we shall be able to identify the trace of the forcing and distinguish the properties of turbulence that are
linked to it from generic ones: forcing-dependent and forcing-independent properties shall indeed be segregated on the grounds that since current is injected at one wall only, the dimensional trace of the forcing is carried by the antisymmetric part of these profiles, while their symmetric part is relatively independent of the forcing (section \ref{sec:wall_scalings}).
Driving the flow in this manner offers a very general way to parametrise a wide 
range of forcing types: any force density $\mathbf f$ applied to the flow can indeed 
be characterised by its intensity and geometry. The former can be measured 
through the current $\rho f/B$ it induces, and the latter by the way in which 
this current distributes in the flow. These in turn determine the intensity of 
the average flow and turbulent fluctuations, which we shall analyse. In 
electrically driven flows,  the total injected current therefore provides a 
measure of the total force applied on the flow. Since the dimensionality of 
the forcing is controlled by the intensity of the externally imposed magnetic 
field, the forcing geometry is precisely controlled along the direction of the magnetic field as well as in the plane across it.
\section{Theory
\label{sec:theory}}
We shall first establish generic scalings for the bulk velocity in MHD channel 
flows driven by injecting electric current at one of the walls (represented in 
figure \ref{fig:3dsketch}).
\subsection{Governing equations and governing parameters}
\label{sec:equations}
In the low-$R\!m$ approximation (\cite{roberts67}), the motion of an electrically conducting fluid (kinematic 
viscosity $\nu$) in an ambient magnetic field $\mathbf B$ induces an electric 
current density $\mathbf J$. In turn, their mutual interaction creates a Lorentz force density $\mathbf J\times \mathbf B$ on the flow, 
which drives the  physical mechanisms analysed in this paper.
The magnetic field associated to the induced current is of the order of 
$\mu\sigma Uh B=R\!m B$ ($\mu$ is the magnetic permeability of the fluid). Since, in the experimental configuration considered 
here, 
the magnetic Reynolds number $R\!m$ remains well below $10^{-1}$, 
the flow-induced component of the magnetic field can be safely neglected in the 
expression of the magnetic field and $\mathbf B$ shall coincide with the 
externally imposed field $B \mathbf e_z$. In 
these conditions,  mechanical and electromagnetic quantities are only coupled 
through the mutual dependence of the velocity and pressure fields $\mathbf u$ 
and $p$ on one side, and the electric current density $\mathbf J$ on the other. The governing equations consist of the Navier-Stokes equations
\begin{equation}
\partial_t \mathbf u+ \mathbf u\cdot\nabla\mathbf u+\frac1\rho\nabla p=\nu\nabla^2 \mathbf u+\frac{B}{\rho}\mathbf J\times\mathbf e_z,
\label{eq:ns}
\end{equation}
Ohm's law
\begin{equation}
\frac1\sigma\mathbf J=-\nabla\phi+B\mathbf u\times \mathbf e_z,
\label{eq:ohm}
\end{equation}
where $\phi$ is the electric potential, and the conservation of mass and charge, respectively:
\begin{eqnarray}
\nabla\cdot\mathbf u &=& 0 \label{eq:contu},\\
\nabla\cdot\mathbf J &=& 0 \label{eq:contj}.
\end{eqnarray}
A Poisson equation for the electric field can be obtained from (\ref{eq:ohm}) and (\ref{eq:contj}). Replacing (\ref{eq:contj}) with it and substituting (\ref{eq:ohm})
into (\ref{eq:ns}) provides a $\mathbf u-p-\phi$ formulation of (\ref{eq:ns})-(\ref{eq:contj}), which is convenient for numerical simulations (\cite{dp12_jfm}). 
With a choice of reference velocity $U$, reference lengths across and along 
the magnetic field $l_\perp$ and $h$, time, pressure, electric current density
 and electric potential can respectively be normalised by $l_\perp/U$, 
$\rho U^2$, $\sigma BU$ and $BUh$. It turns out that the system 
(\ref{eq:ns}-\ref{eq:contj}) is determined by three non-dimensional parameters:
\begin{equation}
H\!a=Bh\sqrt\frac{\sigma}{\rho\nu}, \qquad
N=\frac{\sigma B^2l_\perp}{\rho U}, \qquad
\lambda=\frac{l_\perp}{h}. \qquad
\label{eq:nd_param}
\end{equation}
The square of the Hartmann number $H\!a^2$ expresses the ratio of Lorentz to viscous forces, while the interaction parameter $N$ is a coarse estimate of the ratio of Lorentz to inertial forces. 
The effect of the Lorentz force at low $R\!m$ comes into light through the curl 
of Ohm's law (\ref{eq:ohm}), and by virtue of (\ref{eq:contu}):
\begin{equation}
\nabla\times \mathbf J=\sigma B\partial_z \mathbf u.
\label{eq:curlj}
\end{equation}
%
(\ref{eq:curlj}) expresses that gradients of {horizontal velocity along $\mathbf B$ induce electric eddy currents with a vertical component}.
The associated Lorentz force density $\mathbf F_L=\mathbf J\times\mathbf B$ tends to dampen this velocity gradient. In the absence of a free surface, only its rotational part shall affect the flow, which, by virtue of (\ref{eq:curlj}) can be expressed through 
\begin{equation}
\nabla^2 \mathbf F_L=-{\sigma B^2}\partial^2_{zz} \mathbf u+ \nabla (\sigma B^2 \partial_z u_z).
\label{eq:curlf}
\end{equation}
In configurations where boundary conditions guarantee the existence of the 
inverse Laplacian, \cite{sm82} deduced from this expression that 
at low $R\!m$, the Lorentz force diffuses momentum in the direction of the 
magnetic field over a distance 
$l_z$, in characteristic time $\tau_{2D}(l_z)=\tau_J(l_z/l_\perp)^{2}$. 
Interestingly, this mechanism does not involve the momentum equation 
(\ref{eq:ns}) and thus 
remains valid regardless of the incompressible fluid considered. Based on this 
more precise phenomenology, \cite{sreenivasan02} defined a \emph{true} 
interaction parameter $N_t=N\lambda^{2}$, which represents the ratio of 
the diffusive effect of the Lorentz force to inertial effects more accurately 
than $N$. {In the limit $N_t\rightarrow\infty$, $H\!a\rightarrow\infty$, the Lorentz force dominates, and boundary conditions permitting, the 
flow may then become strictly two-dimensional}. Eddy currents then vanish and 
so does the Lorentz force. A through current $J_z$ can still exist, 
but does not interact with the flow.\\
%
Let us now turn to the 
canonical configuration of flows 
bounded by two electrically insulating walls orthogonal to $\mathbf B$, located 
at $z=0$ and $z=h$ (figure \ref{fig:3dsketch}). 
 The flow is driven by injecting a DC electric current  $I$ at point electrodes embedded in the otherwise electrically insulating bottom wall.  
In this case, the no-slip boundary condition at the wall imposes that viscous 
friction must oppose $\mathbf F_L$ in the Hartmann boundary layers near the 
wall. The balance between these forces determines the Hartmann layer thickness
 as $\delta_H=h/H\!a$. In the absence of inertia, their laminar profile
 is a simple exponential function of the distance to the wall (see for instance \cite{moreau90} for the full theory of these layers). 
%
 In the limit $H\!a\rightarrow\infty$, $N\rightarrow\infty$,  
\cite{sommeria86} showed that for a single point-electrode, the azimuthal 
velocity in the core at $(r,z=\delta_H)$ was linearly dependent on the current $I$ 
injected through the electrode. With a free surface present at $z=h$, this 
translated into
\begin{equation}
U_b (r)=\frac{I_b}{2\pi r \sqrt{\rho\sigma\nu}},
\label{eq:ub}
\end{equation}
where  $(r,z)$ are the cylindrical coordinates {associated with} 
the electrode. 
In a channel, quasi-two-dimensionality in 
the core implies that a second boundary layer, exactly symmetric to the bottom 
one, is present near the top wall. An example of such a configuration, with two 
axisymmetric vortices, is represented in figure \ref{fig:3dsketch} (c). The 
radial currents in the top and bottom Hartmann layers, $I_t$ and $I_b$ 
satisfy $I_t=I_b=I/2$, and so $U_b$ is half of the value found when a free 
surface is present.
%
%
\subsection{Three-dimensional, inertialess flow}
\label{sec:axi_visc}
Since we are interested in three-dimensional flows, which were not considered in these earlier studies, we now turn our attention to cases where either $H\!a$ 
or $N$ is finite. From the curl of (\ref{eq:ns}), finite viscous or inertial 
effects drive a divergent current in the core:
\begin{equation}
\partial_z J_z=-\nabla_\perp\cdot\mathbf J_\perp=\frac\rho{B}(\mathbf u\cdot\nabla\mathbf \omega_z + \mathbf \omega\cdot\nabla{ u_z}) -\frac{\rho\nu}B\nabla^2 \mathbf \omega_z.
\label{eq:j3d}
\end{equation}
Let us first consider the case where $Ha$ is finite but inertia can be neglected ($N\rightarrow\infty$), 
equation (\ref{eq:j3d}) then expresses that viscous 
forces in the core are balanced by a purely rotational Lorentz force driven by 
horizontally divergent  
eddy currents "leaking" into the core (from the centre of the vortex, in case of an axisymmetric vortex). This is represented schematically in figure 
\ref{fig:3dsketch} (a). From (\ref{eq:curlj}), the 
intensity of these currents is proportional to the gradient of velocity in the 
core, which decreases away from the bottom wall. At a distance $l_z^\nu$ from 
the electrode along $\mathbf e_z$, the total 
vertical current injected in the core at $z=0$ is exhausted and the vortex dies 
out. Here $l_z^\nu$ {is found from the third curl of (\ref{eq:ns}), using (\ref{eq:curlf})}:
\begin{equation}
\sigma B^2\partial^2_{zz} \omega_z=\rho\nu\nabla^4 \omega_z.
\label{eq:3dvisc}
\end{equation}
Assuming that in the core, derivatives along $\mathbf e_z$ are of the same order or smaller than radial ones, it follows that:
\begin{equation}
\frac{l_z^{\nu}}{l_\perp}\sim \frac{l_\perp}{h} H\!a=\frac{l_\perp}{\delta_H}. 
\label{eq:lznu}
\end{equation}
%
If $l_z^\nu<h$, the current $I$ injected through the electrode doesn't reach the upper wall but spreads between the core and the bottom boundary layer:
\begin{equation}
I=I_c+I_b.
\label{eq:total_current_b}
\end{equation}

The total current that feeds horizontally divergent currents is 
$I_c\sim 2\pi l_\perp l_z^\nu J_\perp$. Here $J_\perp$ is estimated from 
(\ref{eq:j3d}) as follows: since 
$\nabla^2\sim l_\perp^{-2}(1+\delta_H^2/l_\perp^2)\sim l_\perp^{-2}$, it comes that $J_\perp\sim\rho\nu U_b/(Bl_\perp^2)$ and 
$I_c\sim2\pi l_\perp U_b (\rho\sigma\nu)^{1/2}$. Also, by virtue of (\ref{eq:ub}), 
the current density and the total current in the Hartmann layer express 
respectively as 
$J_b\sim\sigma B U_b$ and $I_b\sim 2\pi l_\perp \delta_H J_b\sim 2\pi l_\perp U_b (\rho\sigma\nu)^{1/2}$. It turns out that $I_b\sim I_c\sim I/2$, so the 
injected current spreads equally between the core and the boundary layer,
 and from (\ref{eq:total_current_b}), the corresponding velocity outside the bottom Hartmann layer is
\begin{equation}
U_b\sim\frac12 \frac{I}{2\pi l_\perp\sqrt{\sigma\rho\nu}}.
\label{eq:ub0_dim}
\end{equation}
This new result expresses that the scaling (\ref{eq:ub}), which was derived for quasi-two-dimensional flows extends to three-dimensional flows where the three-dimensionality originates from viscous effects in the core. Based on this, we shall 
define two distinct Reynolds numbers to characterise respectively the forcing 
and the measured flow intensity:\\
\begin{eqnarray}
Re^0&=&\frac{K\Gamma}{2\nu} \qquad{\rm with}\quad \Gamma=\frac{I}{2\pi\sqrt{\sigma\rho\nu}} ,
\label{eq:reo}\\
Re^b&=&\frac{U_b l_\perp}\nu,
\label{eq:reb} 
\end{eqnarray}
%
where $K$ accounts for the geometry of the current injection pattern. 
For a single vortex, $K=1$ ; in a square vortex array of alternate spin and 
step $L_i$, the velocity induced between two electrodes is $2\Gamma/(L_i/2)$ 
and so $K=4$, as in \cite{kp10_prl}. In regimes where three-dimensionality is 
driven by viscous effects, equation (\ref{eq:ub0_dim}) translates into
%
\begin{equation}
Re^b\simeq C_0 Re^0,
\label{eq:reb_th_0}
\end{equation}
which is of the same form as the scaling for quasi-two-dimensional flows. 
{Both scalings were obtained under the assumption that inertia 
was negligible. Viscous forces were neglected in the bulk to obtain the quasi-two-dimensional scaling, but weren't neglected in (\ref{eq:reb_th_0}). In this sense, 
the latter generalises the former, and they are both part of an \emph{inertialess} regime}.
In (\ref{eq:reb_th_0}) and the remainder of this section, all constants $C$ and 
$D$ with various indices are real scalars of the order of 
unity. The physical mechanism underlying (\ref{eq:lznu}) and (\ref{eq:reb_th_0}) is the same as 
that discovered by \cite{ludford61_jfm} and \cite{hunt68}. It explains the 
presence of a zone of stagnant fluid in regions attached  
to solid obstacles and  extending over a distance of the order of $H\!a$ along $\mathbf B$. Hunt's wake, as it was later called, was found in the numerical simulations of \cite{dp12_jfm} and in the recent experiments of \cite{andreev13_jfm}. \cite{alpher60_rmp}, also reported the existence of a similar wake attached to a conducting strip placed in a free surface channel flow.\\
%
\subsection{Inertial electrically driven flow
\label{sec:axi_inertia}}
\begin{figure}
\centering
\psfrag{Ic}{\textcolor{red}{$I_c$}}
\psfrag{Ic0}{\textcolor{red}{$I_c\simeq0$}}
\psfrag{Ib}{\textcolor{red}{$I_b$}}
\psfrag{It}{\textcolor{red}{$I_t$}}
\psfrag{Itb}{\textcolor{red}{$I_t=I_b$}}
\psfrag{I}{\textcolor{red}{$I$}}
\psfrag{Ub}{\textcolor{blue}{$U_b$}}
\psfrag{Ut}{\textcolor{blue}{$U_t$}}
\psfrag{Utb}{\textcolor{blue}{$U_t=U_b$}}
\psfrag{dh}{{$\delta_H$}}
\psfrag{h}{{$h$}}
\psfrag{lz}{{$l_z$}}
\hspace{-1cm}
\includegraphics[height=0.4\textwidth]{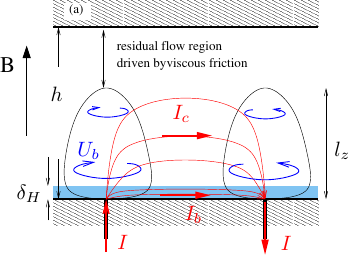}
\includegraphics[height=0.4\textwidth]{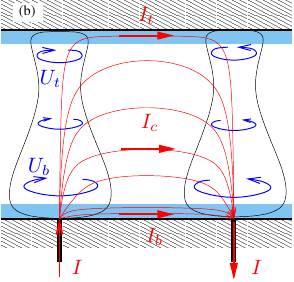}
\includegraphics[height=0.4\textwidth]{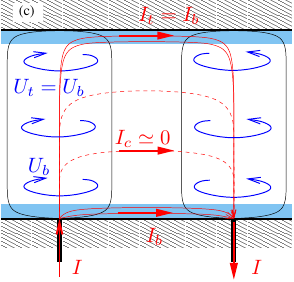}
\hspace{0.5cm}
\includegraphics[height=0.35\textwidth]{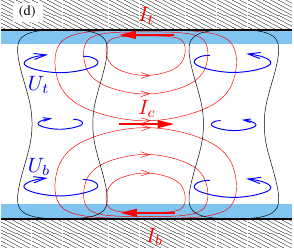}
\caption{Schematic representation of generic flow configurations in 
a channel in an {external} magnetic field. Hartmann layers are represented in light blue, paths of electric current in red, and fluid flow in blue. 
(a) $l_z<<h$: the injected current 
spreads between the Hartmann layer and the core, where the Lorentz force it generates balances either viscous or inertial forces. (b) $l_z\gtrsim h$: same as (a) but part of the injected current flows into the top Hartmann layer. The flow is three-dimensional and influenced by the top wall. (c) $l_z>> h$: the current spreads equally between top and bottom Hartmann layers, with practically no leak into the core: the flow is quasi-two-dimensional. (d) symmetrically three-dimensional structures, not attached to electrodes and where eddy currents in the core are equally sourced from both Hartmann layers. }
\label{fig:3dsketch}
\end{figure}

In turbulent regimes, $N$ is finite and 
inertia may act  
in the core with $u_r\sim u_\theta\sim U_b$ for any given vortex, so that 
$\mathbf u\cdot\nabla\omega \sim {U_b^2/l_\perp}$. By virtue of (\ref{eq:j3d}), 
it must be balanced by the Lorentz force and this pulls some of the current 
injected at the electrode into the bulk of the flow. Then, neglecting viscous 
effects in (\ref{eq:j3d}), and using  (\ref{eq:curlj}), as in section 
\ref{sec:axi_visc}, it comes that the fluid is set in motion up to a distance 
$l_z^{(N)}$ from the wall. Here $l_z^{(N)}$ corresponds to the diffusion length by the Lorentz force in the presence of inertia first introduced by \cite{sm82}:
\begin{equation}
\frac{l_z^{(N)}}{l_\perp}\sim N^{1/2}.
\label{eq:lznonaxi}
\end{equation}
%
Using the same approach as in 
section \ref{sec:axi_visc}, and still considering that $l_z<h$, we note that 
in the limit $H\!a\rightarrow\infty$ and for finite $N$, (\ref{eq:j3d}) implies 
 that the horizontally divergent core current scales as $J_\perp\sim \rho U_b^2l_\perp^{-1}B^{-1}$. {Furthermore, the scaling for the current in the Hartmann layer $I_b\sim2\pi l_\perp U_b (\sigma\rho\nu)^{1/2}$ 
remains 
valid and so from the conservation of current $I=I_c+I_b$}, we arrive at a new scaling for the velocity outside the bottom 
boundary layer:
\begin{equation}
U_{b}\sim\frac1{1+(R\!e^b)^{1/2}}\frac{\Gamma}{l_\perp}.
\label{eq:ubn}
\end{equation}
%
For $Re^b>>1$, expressing $\Gamma$ with (\ref{eq:reo}), the inertial regime is thus characterised by a scaling of the form:%
\begin{equation} 
Re^b\simeq C_b^{(N)} {(Re^0)}^{2/3}.
\label{eq:reb_th_inert}
\end{equation}
For a pair of vortices, velocity and current distributions are qualitatively similar to those represented in figure \ref{fig:3dsketch}, but differ from the viscous case of section \ref{sec:axi_visc} in how they scale with $Re^0$.\\

A somewhat similar distinction between viscous and inertial regimes in electrolytes 
flows driven by  passing an imposed current through an inhomogeneous magnetic 
field was observed experimentally by \cite{duran-matute10_pre}. In this 
case, however, the current was not influenced by the flow so the Lorentz 
force acted as an externally imposed force, and this led to different scalings to those we find here.
In the viscous regime, they found a scaling 
consistent with (\ref{eq:ub}), whereas in our notation, the Reynolds number in 
their inertial regime scaled as 
\begin{equation}
Re^b\sim {(Re^0)}^{1/2}.
\label{eq:duran}
\end{equation}
This scaling reflects a  different forcing mechanism whereby the electric 
current, and therefore the Lorentz force, spread across the whole fluid layer 
instead of being largely confined to the boundary layers, as here at high 
Hartmann numbers. The Lorentz force was then directly balanced by inertial terms. 
%
%
\subsection{Measure of three-dimensionality at the Hartmann wall where no current is injected}
\label{sec:3d_walls}
Let us now consider the influence of the top Hartmann wall present at $z=h$. 
If, as in the previous section, 
$h> l_z$, then only weak residual flows and currents exist in its 
vicinity and the influence of the top wall on the flow can be expected to be 
minimal. In this sense, the top wall is \emph{passive}.
If, on the other hand $h<l_z$, the current injected through an electrode 
mounted at the bottom wall separates in three components instead of two when 
$h>l_z$: the first, $I_b$, flows from the electrode directly into the 
bottom Hartman layer and satisfies 
$U_b\sim I_b/(2\pi l_\perp\sqrt{\sigma\rho\nu} )$ as previously. The second one 
is pulled into the core either by viscous or inertial forces, depending on the 
values of $H\!a$ and $N$. The third one is made of the 
residual current $I_t$ at $z\sim h-\delta_H$.
$I_t$ flows in the top Hartmann layer and in this sense, the top 
wall "cuts" the vortical structure at $z=h$, where a significant 
flow $U_t$ still exists and satisfies 
$U_t\sim I_t/(2\pi l_\perp\sqrt{\sigma\rho\nu})$ by virtue of 
(\ref{eq:curlj}). In this sense, the top wall is \emph{active}.\\
In the inertialess regime, $N\rightarrow\infty$ and (\ref{eq:ns}) expresses that the
 current pulled into the core 
for the Lorentz force to balance viscous forces there is $I_c\sim 2\pi l_\perp \rho\nu h U_b/(Bl_\perp^2)$. Then, since $I=I_b+I_c+I_t$, it comes that the inertialess scaling (\ref{eq:reb_th_0}) found for $Re^b$ in the absence of the top wall remains valid and we arrive at a new scaling for the velocity near to the top wall:
\begin{equation}
U_t\sim U_b\left(1-\frac{h}{l_z^\nu}\right).
\label{eq:utnu}
\end{equation}
{
Alternately, this scaling can also be obtained by noting that $U_t\simeq U_b+h\partial u/\partial z\simeq U_b-h(U_b/l_z^\nu)$.}
Using $Re^t$, a Reynolds number based on the top velocity $U_t$, this yields a 
non-dimensional scaling of the form:
\begin{equation}
Re^t\simeq C_t^{0} Re^0\left(1-D_t^{0}\frac{h^2}{l_\perp^2}{H\!a}^{-1}\right).
\label{eq:ret_th_0}
\end{equation}
The top wall thus affects the distribution of current in the core, mostly in 
its vicinity but not near the bottom wall.\\
The same reasoning applies to the inertial regime ($N$ finite, $H\!a\rightarrow\infty$). This time,
the scaling for the velocity outside the top Hartmann layer becomes
\begin{equation}
U_t\sim U_b\left(1-\frac{h}{l_z^{(N)}}\right).
\label{eq:ut_walls}
\end{equation}
Three cases can be distinguished for the expression of $I_c$, depending on $l_z^{(N)}/h$.
\begin{itemize}
\item[(1)] For $N_t<1$ ($l_z^{(N)}< h$), the top wall is passive: the scaling of section \ref{sec:axi_inertia} holds and $I_c\sim2\pi l_\perp \rho U_b^2/(B l_\perp) l_z^{(N)}$.
\item[(2)] For $N_t\gtrsim1$ ($l_z^{(N)}\gtrsim h$), noticeable three-dimensionality is present in the core.  Most of the eddy currents recirculate in the core. Only a small portion of these currents passes through the top Hartmann layer, without significantly affecting the core current. Then $I_c$ can be expected to still scale as $I_c\sim2\pi l_\perp \rho U_b^2/(B l_\perp) l_z^{(N)}$.
\item[(3)] In the limit $N_t>>1$ ($l_z^{(N)}>>h$), the flow is close to quasi-two-dimensionality and most eddy currents recirculate through both Hartmann layers. The current in the core scales as $I_c\sim2\pi l_\perp \rho U_b^2/(B l_\perp) h$, which is typically $H\!a$ times smaller than the currents in the Hartmann layers $I_b$ and $I_t$. 
\end{itemize}
All three cases can be reconciled into one in writing 
$I_c\sim2\pi l_\perp \rho U_b^2/(B l_\perp) l_z^{(N)}f(h/l_z^{(N)})$. 
The correction $f(h/l_z^{(N)})$ then varies from a constant value for $l_z^{(N)}<h$ (case (1)), then exhibits dependence on $N_t$ for $N_t\gtrsim1$ (case (2)), and tends to a function of order $N_t^{-1/2}$ in the limit $l_z/h\rightarrow\infty$ (case (3)). The scaling for $Re^b$ follows from the global current conservation, $I=I_b+I_t+I_c$:
\begin{equation}
\left(2 (Re^b)^{-1/2} -(Re^b)^{-1/2} N_t^{-1/2} + f(N_t)\right) (Re^b)^{3/2}\simeq (C_b^{(N)})^{3/2} Re^0.
\label{eq:reb_wall}
\end{equation}
In the limit of large $N_t$ ($l_z^{(N)}>>h$), the first term in the bracket dominates and the law for quasi-two-dimensional flows is recovered. The middle term represents a correction to the current in the top layer accounting for the current lost to the core. The last term comes from the current pulled by inertial effects into 
the core. For $N_t\gtrsim1$ ($l_z^{(N)}\gtrsim h$), it approaches a constant and becomes larger than the other two. This case shall be the most interesting one, since it combines inertia-induced three-dimensionality with an active influence of the top wall. In this limit using (\ref{eq:ut_walls}) for $U_t$, $Re^b$ and $Re^t$ can be expressed as:
\begin{eqnarray}
Re^b&\simeq& C_b^{(N)} {(Re^0)}^{2/3} f(N_t)^{-2/3},
\label{eq:reb_th_inert_walls}\\
Re^t&\simeq& C_t^{(N)} {(Re^0)}^{2/3}f(N_t)^{-2/3} \left(1-D_t^{(N)}{N_t^{-1/2}}\right).
\label{eq:ret_th_inert}
\end{eqnarray}
Note that both (\ref{eq:ret_th_0}) and (\ref{eq:ret_th_inert}) have the same form 
as their counterparts in the vicinity of the wall where current is injected but 
for a coefficient smaller than unity. This coefficient therefore gives a global 
measure of the amount of three-dimensionality across the fluid layer. 

%
%
\subsection{Symmetric \emph{vs.} antisymmetric three-dimensionality
\label{sec:s_as_3d}}
Equation (\ref{eq:j3d}) imposes that inertial or viscous effects in the 
core pull a horizontally divergent current there and induce a variation 
of vertical current. It imposes, however, no constraint on the direction in 
which this current travels along $\mathbf B$. This 
shall be decided by the boundary conditions of the problem. In the examples of figures 
 \ref{fig:3dsketch} (a), (b) and (c), the current enters and leaves the domain 
at two electrodes located at the bottom wall. The topology of the electric 
current lines must ensure that the circuit is closed whilst satisfying this 
topological constraint and this forces the current to flow as sketched. 
A structure not attached to an injection electrode may, by contrast, not 
necessarily exhibit the asymmetry imposed by the location of the electrode and the current that 
loops in the core may be sourced from either Hartmann layers. In this case, 
\cite{psm00_jfm}'s asymptotic analysis of a symmetric channel (with no 
injection electrode) 
shows that when $N$ decreases from infinity, the first form of 
three-dimensionality encountered is symmetric, with a velocity profile  
that is quadratic in $z$. The underlying mechanism was indeed the 
same as in sections \ref{sec:axi_inertia} and 
\ref{sec:3d_walls}, except that eddy currents feeding the core current took 
source equally in the top and bottom Hartmann layers, as sketched in figure 
\ref{fig:3dsketch} (d).\\
Such \emph{symmetric} three-dimensionality may still occur in a channel where 
the flow is driven by injecting current at one wall only. In turbulent regimes 
indeed, energy is transferred from the mean flow, made of vortices attached to 
the injection electrodes (characterised by $l_\perp^{\rm mean}$ and $U_b^{\rm mean}$), to transient structures, not attached to electrodes. The transfer takes place over a structure turnover time $\tau_u=l_\perp^{\rm mean}/U_b^{\rm mean}$. After typically 
$\tau_{2D}(l_\perp^{\rm mean})=(h/l_\perp^{\rm mean})^2\rho/(\sigma B^2)$, 
momentum diffusion by the Lorentz force has erased the asymmetry inherited from
the influence of the electrodes on the the mean flow and \cite{psm00_jfm}'s theory 
becomes relevant. 
Depending on their sizes, such structures can be either two or 
three-dimensional. 
The possible existence of symmetrically three-dimensional 
structures can be seen by noting that for a structure of size $l_\perp$ 
to be three-dimensional, its turnover time must satisfy 
$\tau_{2D}(l_\perp)>l_\perp/U(l_\perp)$ {(or equivalently $N_t(l_\perp)<1$). 
For it to be symmetric, on the other hand, the energy transfer time from the 
mean flow to it, which is at least $l_\perp^{\rm mean}/U_b^{\rm mean}$, must 
be larger than $\tau_{2D}(l_\perp^{\rm mean})$ (or $N_t(l_\perp^{\rm mean})>1$).
For $l_\perp<l_\perp^{\rm mean}$,  $\tau_{2D}(l_\perp^{\rm mean})<\tau_{2D}(l_\perp)$, and so depending on the variations of $U(l_\perp)$, both conditions may 
be satisfied (for instance if $U(l_\perp)\sim U_b^{\rm mean}$), and symmetric three-dimensionality may be present in turbulent fluctuations, even though the 
mean flow may be asymmetric if $N_t(l_\perp)<1<N_t(l_\perp^{\rm mean})$}. A consequence is that forcing the flow through electrodes embedded in one wall can still lead to turbulence that is either 
quasi-two-dimensional or three-dimensional and not influenced by the dimensionality of the forcing.
%
\section{Measure of flow dimensionality through scaling laws for the velocity
\label{sec:visc-inertia}}
\subsection{Experimental measurements of $U_b$ and $U_t$ 
\label{sec:u_estimates}}
%
We shall now experimentally characterise the origin of three-dimensionality, 
 using the scalings from  section \ref{sec:theory}. These involved bulk 
velocities near top and bottom Hartmann walls, and the injected current per 
electrode $I$. Velocities cannot be directly measured 
in the experiment, but can be estimated using (\ref{eq:u_exp_j}), from 
measurements of $\nabla_\perp\phi$ at the Hartmann walls (see section \ref{sec:m_tech}).
Hence, velocities $U_b$ and $U_t$, which 
appeared in the scalings of section \ref{sec:theory} shall be estimated as
%
\begin{eqnarray}
U_b=B^{-1}\overline{\langle|\nabla\phi_b|\rangle},\\
U_t=B^{-1}\overline{\langle|\nabla\phi_t|\rangle},
\label{eq:u_exp}
\end{eqnarray}
%
where the over-bar indicates spatial averaging in the central square region (see figure \ref{fig:design}), $\langle\cdot\rangle$ stands for time-averaging,
$\phi_b$ and $\phi_t$ are electric potentials measured at the bottom and top 
wall respectively.
Reynolds numbers $Re^b$ and $Re^t$ were built using $l_\perp=L_i$ as reference
length. We shall also extract scalings for their counterparts $Re^{b\prime}$ 
and $Re^{t\prime}$, built on the RMS of velocity fluctuations near the bottom 
and top walls, estimated as
%
\begin{eqnarray}
U_b^\prime=B^{-1}\overline{\langle|\nabla\phi_b-\langle\nabla\phi_b\rangle|^2\rangle^{1/2}},\\
U_t^\prime=B^{-1}\overline{\langle|\nabla\phi_t-\langle\nabla\phi_t\rangle|^2\rangle^{1/2}}.
\label{eq:up_exp}
\end{eqnarray}
For a spacing between probes of 2.5 mm in (\ref{eq:gphi_discrete}), instantaneous velocities can be measured to a precision of 1.6 mm/s (at $B=0.5$ T/ $H\!a=1822.2$) down to 0.16  mm/s (at $B=5$ T/ $H\!a=18222$). In the turbulent regimes considered in this paper, this provides a relative precision of at worse 5\% to 0.5\%. These figures are considerably improved for averaged and RMS quantities, which are calculated over time series of typically $10^5$ samples.
%
%
%
\subsection{Inertialess vs inertial three-dimensionality
\label{sec:ub_scalings}}
\subsubsection{Average flow}
\begin{figure}
\centering
\includegraphics[width=0.8\textwidth]{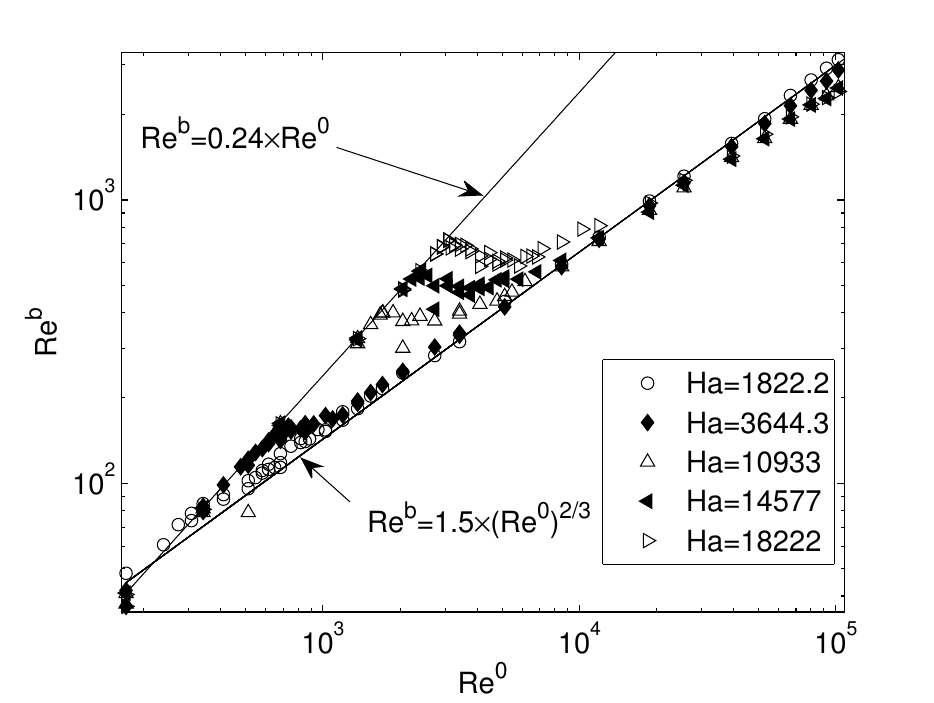}
\includegraphics[width=0.8\textwidth]{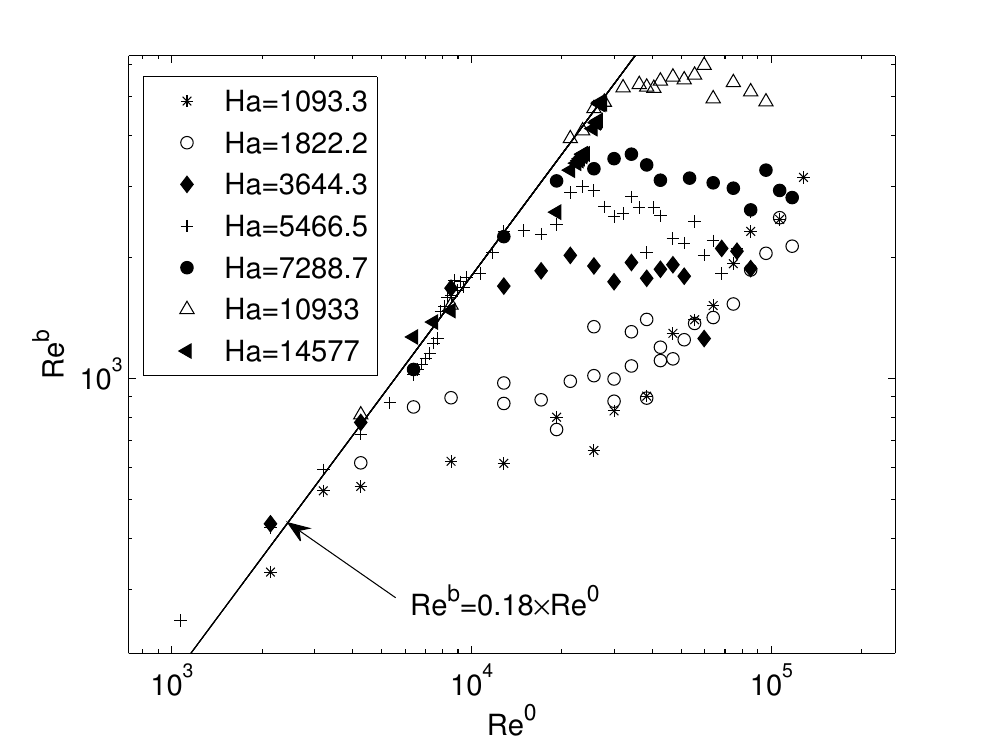}
\caption{Graphs of $Re^b$ \emph{vs.} $Re^0$ representing the average flow near 
the bottom wall for measurements spanning the whole range of parameters 
accessible in the experiment. 
Solid lines represent scaling laws for the inertialess regime (of the form 
(\ref{eq:reb_th_0})) and purely inertial regime (\ref{eq:reb_th_inert}). Top: $\lambda_i=0.1$, bottom: $\lambda_i=0.3$}
\label{fig:reb}
\end{figure}

Figure \ref{fig:reb} shows graphs of $Re^b$ against $Re^0$ over the full
range of parameters spanned in our measurements for $\lambda_i=0.1$ and 
$\lambda_i=0.3$. For any fixed value of $H\!a$, two regimes clearly stand out and data obtained at different values of $H\!a$ all closely collapse into two single curves respectively for lower and higher values of $Re^0$. For 
$\lambda_i=0.1$, these curves follow the scalings
\begin{eqnarray}
Re^b&\simeq&0.24Re^0 \label{eq:reb_exp} \text{ at small } Re^0,\\
Re^b&\simeq&1.5(Re^0)^{2/3} \label{eq:reb_exp_inert} \text{at large } Re^0.
\end{eqnarray}
The scaling at low $Re^0$ is found exclusively in the steady state, where vortices are mostly axisymmetric, and therefore subject to little inertia. It matches
the scaling which characterises the inertialess regime (\ref{eq:reb_th_0}) with $C_0\simeq0.24$, 
which suggests that no three-dimensionality driven 
by inertia is present there. Furthermore, from 
previous observations of flow patterns (\cite{kp10_prl}), the 
reason why this scaling holds at high $H\!a$ is rather that the 
flow is indeed close to quasi two-dimensionality (typically $H\!a>7500$ for 
$\lambda_i=0.1$ and $H\!a>4000$ for $\lambda_i=0.3$). At low $H\!a$, on the other 
hand, three-dimensionality was visible in the base flow, under the form of 
differential rotation affecting the base vortices. In this case, the validity 
of (\ref{eq:reb_exp}) indicates that this effect is driven by viscous 
friction.\\
When the flow becomes unsteady, 
inertia triggers a transition away from the inertialess regime (the conditions 
of this transition where analysed by \cite{kp10_prl} and earlier by \cite{sommeria88} in the quasi-two dimensional regime.). Viscous effects are 
however still present and a purely \emph{inertial} regime, in the sense that 
inertia-driven three-dimensionality dominates, is only reached asymptotically 
at high $Re^0$, when the flow is turbulent. $Re^b$ then 
tends to (\ref{eq:reb_exp_inert}), which matches 
(\ref{eq:reb_th_inert}) with $C_b^{(N)}=1.5$. 
For higher $H\!a$, the asymptotic value of $Re^b$ consistently 
stands slightly under that of (\ref{eq:reb_exp_inert}), as predicted theoretically in (\ref{eq:reb_th_inert_walls}). The correction factor due to the influence of the top wall is however weak, of the order of $f(N_t)^{-2/3}\sim N_t^{0.08 \pm 0.01}$.
Overall, the validity of generic scaling (\ref{eq:reb_th_inert}) for the 
average  flow is verified over a wide range both of $Re^0$ 
and $H\!a$.\\

\begin{figure}
\centering
\includegraphics[width=0.8\textwidth]{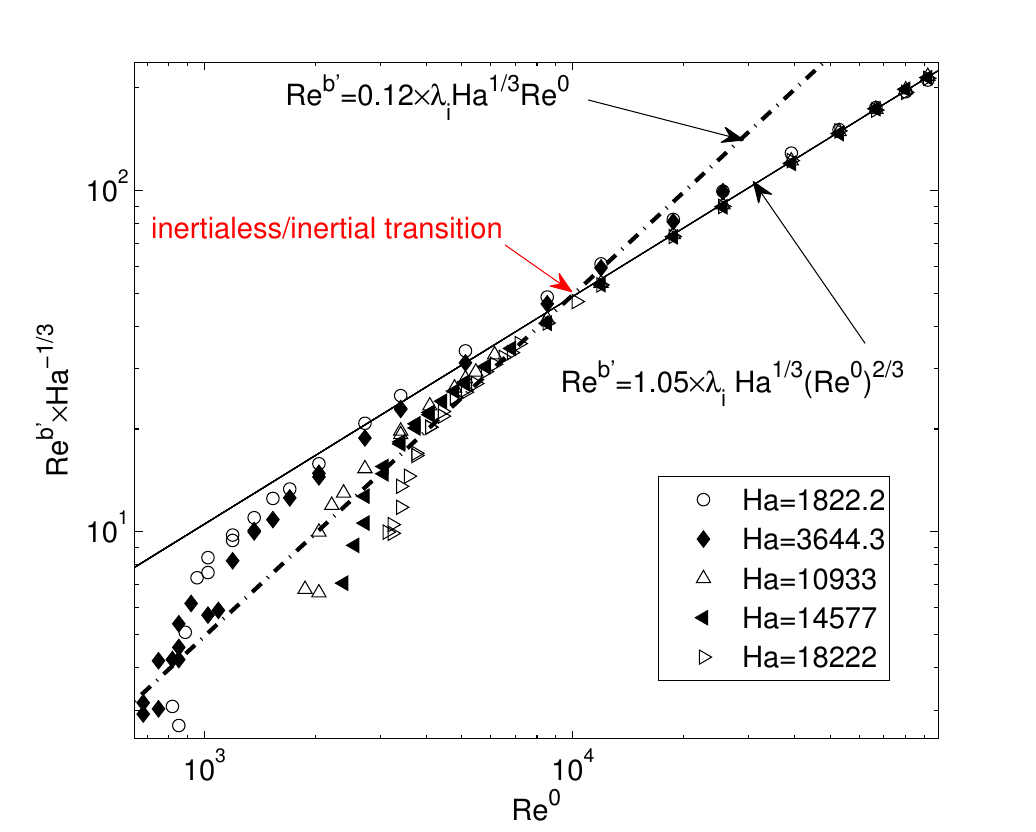}
\includegraphics[width=0.8\textwidth]{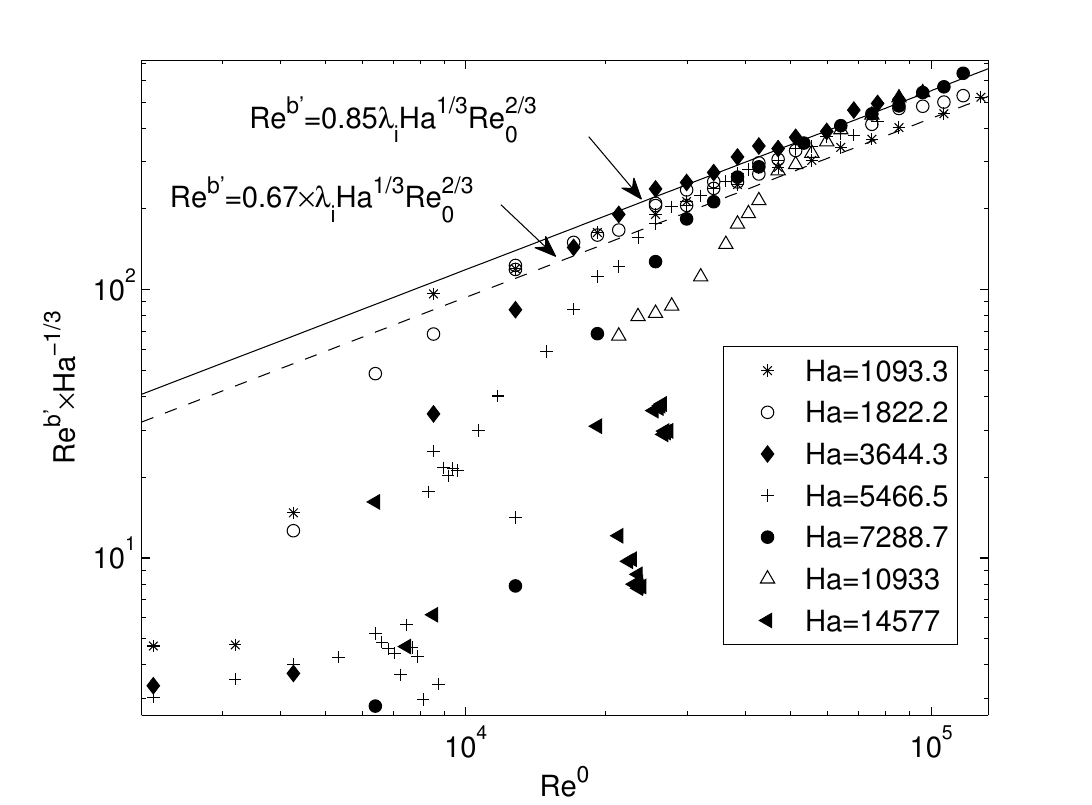}
\caption{Graphs of $Re^{b\prime} H\!a^{-1/3}$ \emph{vs.} $Re^0$, representing the 
RMS of velocity fluctuations near the bottom wall, 
for $\lambda_i=0.1$ (top) and $\lambda_i=0.3$ (bottom). Solid and dashed lines 
represent scaling laws for the purely inertial regime (\ref{eq:reb_th_inert}).}
\label{fig:rebp}
\end{figure}

We were also able to verify that (\ref{eq:reb_th_0}) still holds when the 
forcing scale $\lambda_i$ is varied, even though 
$C_0$ 
is slightly smaller for $\lambda_i=0.3$ than for 
$\lambda_i=0.1$ ( $C_0\simeq0.18$ \emph{vs.} $C_0\simeq0.24$). This point 
is mostly experiment-specific, as larger vortices 
are indeed more sensitive to friction from the side walls because a 
larger proportion of the lattice is in direct contact with them. 
Note that the values of $Re^0$ reached in the experiment for $\lambda_i=0.3$ 
are not large enough for the average flow to reach a fully developed inertial 
regime so we could not verify whether (\ref{eq:reb_th_inert}) was independent of 
$\lambda_i$ for the average flow.\\
%
{\subsubsection{Turbulent fluctuations}
For $\lambda_i=0.1$, fluctuations $U_b^\prime$ also exhibit two distinct 
regimes at low and high $Re$. In the limit $Re^0\rightarrow\infty$, they 
follow a unique law of the form 
$Re^{b\prime}\simeq C^\prime_b\lambda_i(Re^0)^{2/3}H\!a^{1/3}$, with 
$C^\prime_b\simeq1.05$. This law is satisfied to a remarkable precision as soon 
as the flow is 
sufficiently turbulent (see figure \ref{fig:rebp}, top). Experimental data 
shows the scaling $\lambda_i H\!a^{1/3}$ to represent the ratio 
of velocity fluctuations relative to the mean flow in the limit of large Reynolds numbers (see \ref{sec:remarks}\ref{rem:fluc}).
}
Beyond the prefactor in $\lambda_i Ha^{1/3}$, it remains that $Re^{b\prime}\sim (Re^0)^{2/3}$, which is of the form of (\ref{eq:reb_th_inert}), and therefore indicates that fluctuating structures are three-dimensional as a result of 
the inertial mechanism identified in section \ref{sec:axi_inertia}. As for the average flow, the validity of (\ref{eq:reb_th_inert}) is verified over a wide range of $Re^0$ and $H\!a$ for turbulent fluctuations too.\\
{At lower values of $Re$, and for $H\!a=18222$ only, $Re^{b\prime}$ scales as $Re^{b\prime}\simeq0.12\lambda_iH\!a^{1/3}Re^0$, which is of the form (\ref{eq:reb_th_0}) expected in the inertialess limit. (\ref{eq:reb_th_0}) and (\ref{eq:reb_th_inert}) intersect at $Re^{b\prime}\simeq1.27\times10^3$. \emph{This point marks a form of transition between the inertialess regime 
and the regime of inertia-driven three-dimensionality}. This transition occurs at a rather high value of the true interaction parameter $N_t^\prime$ based on $U_b^\prime$, of the order of 26. Nevertheless a more precise estimate of this value, and a confirmation of its validity in the limit $Ha\rightarrow\infty$ would require further experiments at significantly higher magnetic fields.  
For $H\!a<18222$, the inertialess limit is not well achieved and this transition is therefore not as clear.
Remarkably, since the 
inertialess regime for the average flow is only reached in the steady state, 
the inertialess regime for turbulent fluctuations 
corresponds to a state where inertia-driven three-dimensionality is present in 
the base flow but not in the turbulent fluctuations.}\\ 
(\ref{eq:reb_th_inert}) further remains valid for fluctuations when the 
injection scale varies 
albeit with a slightly lower 
value of constant $C^\prime_b$  (0.85, \emph{vs.} 1.05, see figure \ref{fig:rebp}) due to the greater influence of the side walls at higher $\lambda_i$. 
Fluctuations at $\lambda_i=0.3$ further reveal how these 
scalings degenerate at low $H\!a$: the asymptotic law (in the limit 
$Re^0\rightarrow\infty)$ stands a little lower than for higher $H\!a$ 
($C^\prime_b=0.67$). This value of $C^\prime_b$ reflects a non asymptotic 
dependence on $H\!a$:  
the scaling $U_b^\prime/U_b\sim \lambda_i H\!a^{1/3}$ is indeed only valid in the limit 
$H\!a\rightarrow\infty$ but not in the limit $H\!a\rightarrow0$, where non-MHD 
turbulence still sustains strong fluctuations. 
%
\subsection{Measure of three-dimensionality at the wall where no forcing is applied
\label{sec:ut_scalings}}

\begin{figure}
\centering
\includegraphics[width=0.8\textwidth]{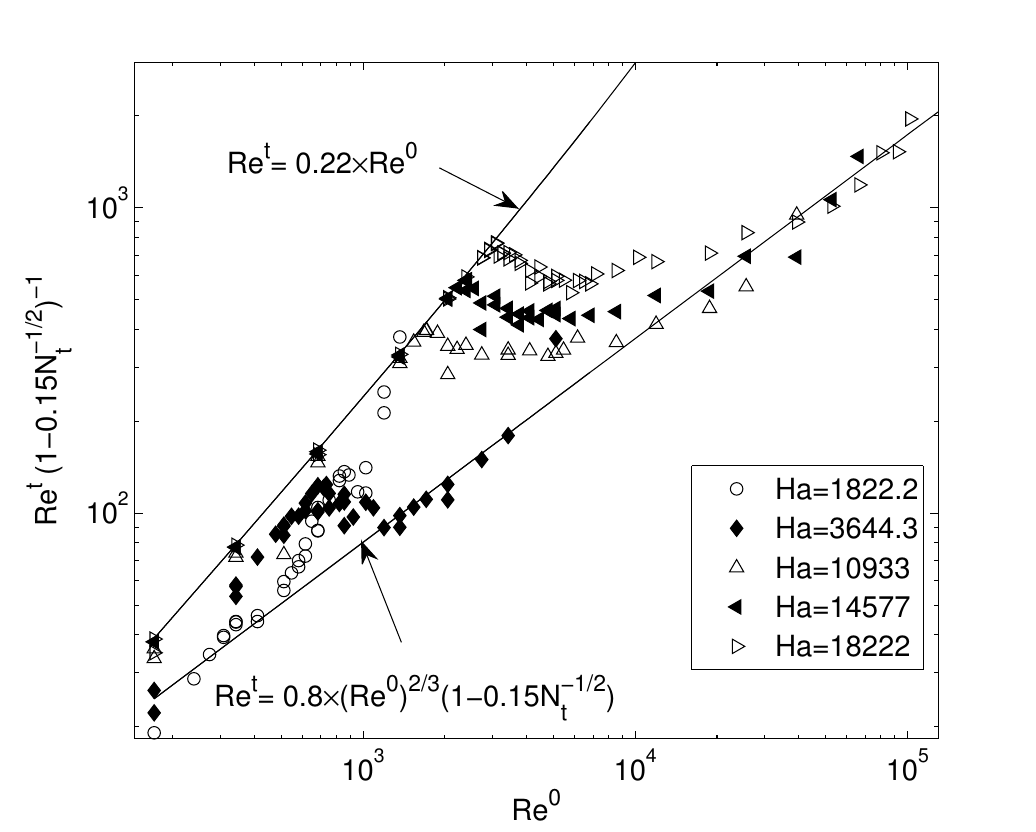}
\includegraphics[width=0.8\textwidth]{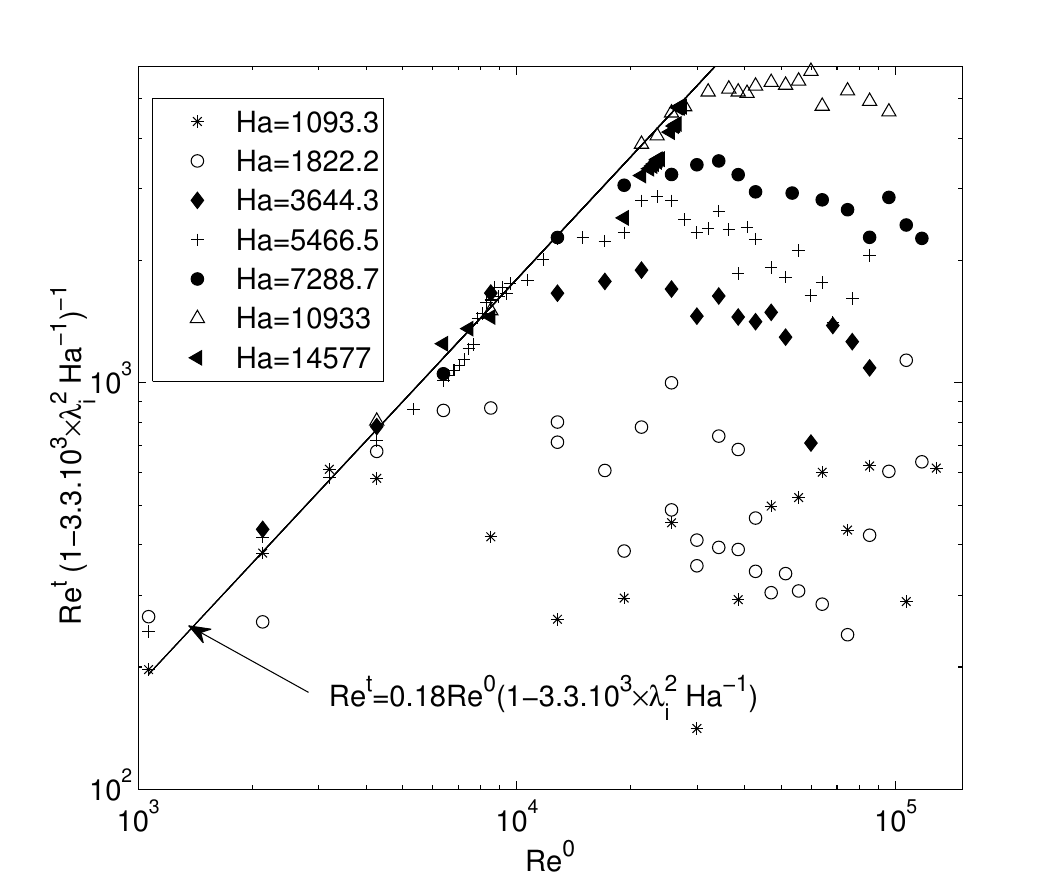}
\caption{Graphs of $Re^t$ \emph{vs.} $Re^0$, 
for $\lambda_i=0.1$ (top) and
 $\lambda_i=0.3$ (bottom). Top: curves are scaled by 
$(1-0.15\times N_t^{-1/2})^{-1}$ to show that all curves collapse into a scaling 
of the form (\ref{eq:ret_th_inert}) in the high $Re^0$ range. A similar collapse
 happens at low  $Re^0$ for scaling (\ref{eq:ret_th_0}) (not represented here for concision). Bottom:  curves are scaled by 
$(1-3.3\times10^3\times\lambda_i^2 H\!a^{-1})^{-1}$ to show that all curves collapse 
into a scaling of the form (\ref{eq:ret_th_0}). Values of $Re^0$ accessible in 
the experiment were not large enough to verify inertial scaling 
(\ref{eq:ret_th_inert}) at $\lambda_i=0.3$.}
\label{fig:ret}
\end{figure}

\begin{figure}
\centering
\includegraphics[width=0.8\textwidth]{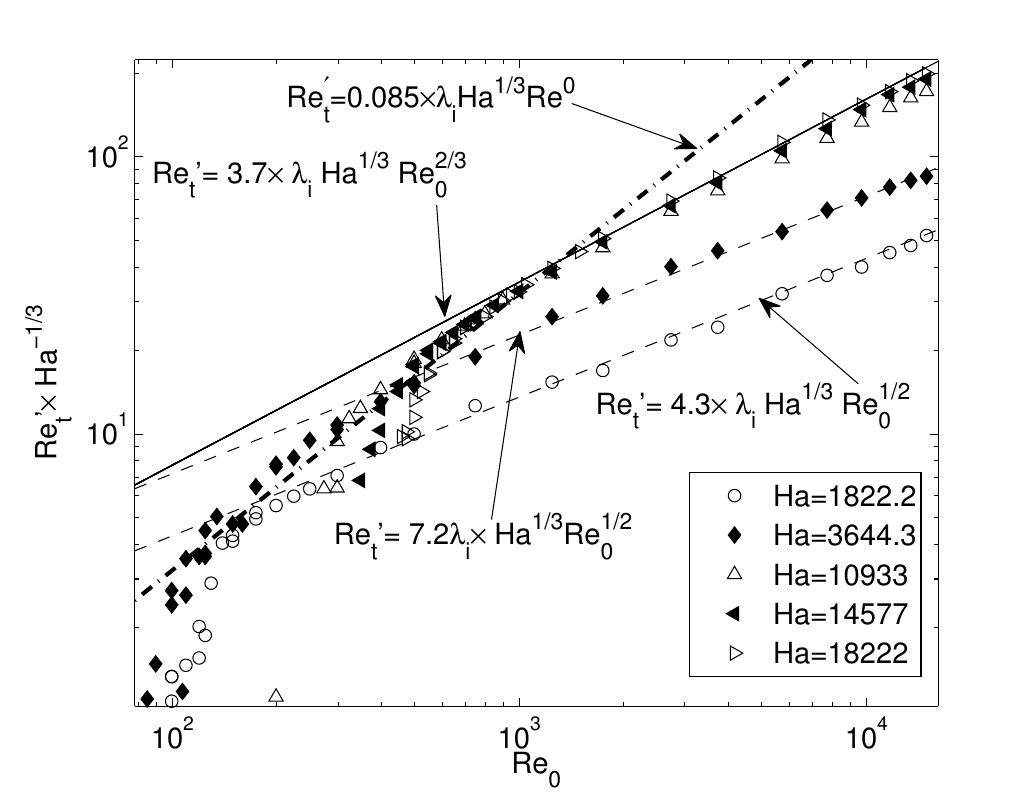}
\includegraphics[width=0.8\textwidth]{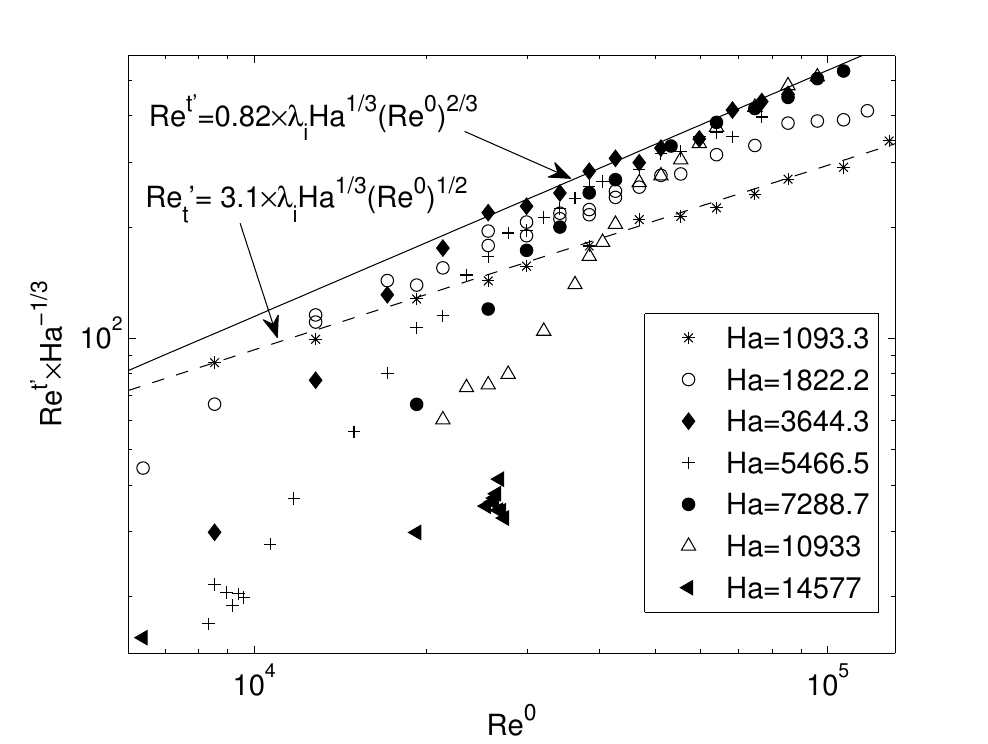}
\caption{Graphs of $Re^{t\prime} H\!a^{-1/3}$ \emph{vs.} $Re^0$, 
for $\lambda_i=0.1$ (top) and $\lambda_i=0.3$ (bottom). Scalings laws corresponding to the purely inertial regime are represented by solid and dashed-dotted lines when the top wall influences the flow ($l_z^{\prime(N)}\gtrsim h$ or $l_z^{\prime(N)}>>h$, scalings (\ref{eq:ret_th_0}) and  (\ref{eq:ret_th_inert})) and by dashed lines when it does not ($l_z^{\prime(N)}< h$, scaling of the form (\ref{eq:duran})).}
\label{fig:retp}
\end{figure}

Now that regimes of quasi-two dimensionality, inertialess and purely inertial 
three dimensionality are identified, we shall examine the 
flow in the vicinity of the upper Hartmann wall to obtain a first quantification of three-dimensionality within these regimes.
This shall be achieved by seeking the 
conditions in which the current injected at the bottom Hartmann wall feeds into 
the Hartmann layers near the top walls (when $l_z\gtrsim h$ and 
$l_z>> h$ according to the theory from section \ref{sec:3d_walls}), and by quantifying $l_z$ (section \ref{sec:wall_scalings}).\\
{\subsubsection{Average flow}}
The variations of $Re^t$ with $Re^0$ are shown in figure \ref{fig:ret} for $\lambda_i=0.1$ and $\lambda_i=0.3$. Their general aspect is somewhat similar to 
those of $Re^b$ with $Re^0$ (figure \ref{fig:reb}), with a linear law  
in the inertialess regime (for $Re^0<Re^0_I$), and a scaling of the form 
$Re^t\sim (Re^0)^{2/3}$ in the purely inertial regime (for $Re^0$ sufficiently 
large). There are, however, two major differences. Firstly, in 
the inertialess regime, a dependence on $H\!a$ of the form (\ref{eq:ret_th_0}) 
is present (with $C_t^0=0.22$ \emph{vs.} $C_b^0=0.24$ for $\lambda_i=0.1$, and $C_t^0=0.18\simeq C_b^0$ for $\lambda_i=0.3$). The graph for $\lambda_i=0.3$ was scaled to find an estimate for the second constant in (\ref{eq:ret_th_0}), 
$D_t^{(H\!a)}\simeq3.10^3$. 
The fact that flow intensities near top and bottom walls are close at high $H\!a$ reflects that the average flow becomes quasi-two dimensional in the limit $H\!a\rightarrow\infty$. At lower $H\!a$, the 
validity of (\ref{eq:ret_th_0}) indicates that the average flow looses intensity away from 
the bottom wall under the effect of viscous friction. The 
second important difference between the variations of $Re^t$ and $Re^b$ is 
visible on the graph for $\lambda_i=0.1$: in the inertial regime too, $Re^t$ 
depends on $H\!a$, but this time according to 
scaling (\ref{eq:ret_th_inert}), which incorporates a correction for inertial effects in the presence of the top wall. 
Together with the validity of (\ref{eq:ret_th_0}) in the range of lower $Re^0$, 
This suggests that the upper wall actively 
influences the flow according to the mechanism outlined in section 
\ref{sec:3d_walls}, where part of the current injected near the bottom wall was 
concentrated in the top Hartmann layer.\\
{\subsubsection{Turbulent fluctuations}
As for the average flow, the variations of the intensity of flow fluctuations 
near the top wall with $Re^0$, reported in figure \ref{fig:retp}, resemble 
their counterpart near the bottom wall but 
exhibit a non-asymptotic behaviour at low $H\!a$. In the limit of large $H\!a$,
$Re^{t\prime}$ clearly follows 1) an inertialess regime at intermediate $Re^0$ and 2) an inertia-driven three-dimensional one in the limit $Re^0\rightarrow\infty$. 
These regimes are respectively characterised by scalings of the form 
$Re^{t\prime}\simeq C_t^{0\prime} \lambda_i Re^0 H\!a^{1/3}$ and  
$Re^{t\prime}\simeq C_t^\prime \lambda_i (Re^0)^{2/3} H\!a^{1/3}$, which only 
differ from the scalings for $Re^{b\prime}$ through slightly lower values of 
$C_t^{0\prime}$ (0.085 \emph{vs.} $C_b^{0\prime}=0.12$) and $C_t^\prime$ (0.88 \emph{vs.} $C_b^\prime=1.05$). The slight differences in the first set of 
constants confirm that three-dimensionality driven by viscous effects is 
present in the inertialess regime.}\\
The most remarkable property of fluctuations near the top plate is that unlike for the base flow, the 
scaling exponent of $Re^0$ in the law $Re^{t\prime}(Re^0)$ is different at low 
$H\!a$ than in the limit of high $H\!a$. At low $H\!a$, it is indeed of the 
form $Re^t\simeq C_t^\prime(H\!a)\lambda_i H\!a^{1/3}(Re^0)^{1/2}$. This 
phenomenon 
is clearly identifiable over one to two decades of $Re^0$ at $\lambda_i=0.1$ 
for $H\!a\leq3644.3$, and at $\lambda_i=0.3$ 
for $H\!a<1822.2$. The corresponding flow fluctuations are of much lower 
intensity than predicted by any of the scenarii of section \ref{sec:3d_walls} 
involving an active influence of the top wall on the flow, so one can infer 
that the electrically driven flow does not reach the upper wall (\emph{i.e.} 
$l_z^{\prime(N)}<h$, where $l_z^{\prime(N)}$ is built on $U_b^\prime$ instead of $U_b$). Instead, the exponent 1/2 of $Re^{b\prime}$ 
is the same as in scaling (\ref{eq:duran}), found by \cite{duran-matute10_pre} in a non-MHD inertial regime. It can 
be understood as follows: for the lowest values of $H\!a$, the flow near the 
bottom wall is entrained 
over a height $l_z^{(N)}<h$ (given by (\ref{eq:lznonaxi})) and so 
the scaling law (\ref{eq:reb_th_0}) for $Re^b$ still holds. Fluctuations, by 
contrast, are not directly driven by current injection but receive energy from 
the average flow by inertial transfer. Furthermore, the non-asymptotic behaviour of $Re^{b\prime}$, visible at low $H\!a$ in figure \ref{fig:rebp} (bottom), 
suggests that for such structures, the core current density becomes comparable to that 
through the bottom Hartmann layer (This electric "leak" from the bottom 
Hartmann layer explains the lower intensity of the fluctuations). Consequently, fluctuations are powered by inertial transfer  
over the whole height $l_z^{\prime(N)}$ of the fluid layer where current passes. The 
Lorentz force balances inertia there, and is of order $\sigma B^2 U_c^\prime$. 
This phenomenology was noted in section \ref{sec:ub_scalings} to translate into a 
fluctuating velocity in the core $U_c^\prime$ 
scaling as $(Re^0)^{1/2}$. The flow in the top region $l_z^{\prime(N)}\leq z\leq h$, is thus not connected to the flow in the region $0\leq z\leq l_z^{\prime(N)}$ by 
eddy currents. Instead, it is entrained by viscous friction acting though 
a fluid thickness of $h-l_z^{\prime(N)}$ (figure \ref{fig:3dsketch} (a)). $U_t^\prime$ is therefore 
proportional to $U_c^\prime$ and decreases with $h-l_z^{\prime(N)}$. This explains that $Re^{t\prime}\sim (Re^0)^{1/2}$. Since the thickness of the top region $h-l_z^{\prime(N)}$ decreases with $H\!a$, it also explains that at low $H\!a$, velocity fluctuations decrease faster when $H\!a$ decreases, than at high $H\!a$, where they scale as $H\!a^{1/3}$ (see \ref{sec:remarks}\ref{rem:ept}).
%

%
%
\section{Symmetric, asymmetric three-dimensionalities and experimental measure of the Lorentz force diffusion length 
\label{sec:wall_scalings}}
From the characterisation of the flow near the wall where no forcing was applied, three-dimensionality was measured and its driving mechanisms identified. We shall now inspect electric potentials along the side walls to determine 
$l_z^{(N)}$ and $l_z^{\prime(N)}$, where, $l_z^{(N)}$ and $l_z^{\prime(N)}$ are respectively built on average and fluctuating quantities. This shall lead us to examine the trace of the forcing in the bulk of the flow, by tracking symmetric and antisymmetric three-dimensionality.

\subsection{Interpretation of electric potential profiles along Shercliff walls}
Unlike Hartmann layers, Shercliff layers, which develop along walls parallel to 
the magnetic field, do not possess a property providing 
 a simple relation between electric potential at the side wall $\phi_S$ and 
velocities 
in the core (such as (\ref{eq:u_exp_j})) and so there is no simple way to deduct 
velocity variations along $\mathbf B$ from time-series of non-intrusive 
measurements of electric potential at the wall. Furthermore, Shercliff layers are 
intrinsically three-dimensional because they result from a balance between 
the Lorentz force and horizontal viscous friction, which, as in the 
mechanisms responsible for three-dimensionality in the core, draws a horizontal 
current from the core into them. Nevertheless, if three-dimensionality is present in the 
core, then the current drawn from it into the Shercliff layers depends on $z$. 
More specifically, the 
\emph{asymmetric} three-dimensionality in the core described in section 
\ref{sec:s_as_3d} translates into an antisymmetric component of the 
profile of $\phi_S(z)$ of order at least $z^3$ (see \ref{sec:remarks}\ref{rem:shercl}). 
Similarly, \emph{symmetric} 
three-dimensionality in the core leads to a symmetric deformation of 
$\phi_S(z)$. The reverse, however isn't true, since even in quasi-two 
dimensional flows, intrinsic three-dimensionality within the Shercliff 
layers induces a symmetric three-dimensional component in $\phi_S(z)$ (see for 
example, \cite{moreau90}). 
Nevertheless, since this deformation is driven by linear viscous friction, it 
must depend linearly on the flow intensity, as measured by $U_b$. 
Consequently, symmetric three-dimensionality in the core can 
still be detected by tracking nonlinear deformations of the symmetric part of 
{the average of} $\phi_S(z)/(BU_bL_i)$, and of {the RMS of} $\phi_S^\prime(z)/(BU_b^\prime L_i)$  with $Re^0$. The distinction between symmetric and asymmetric three dimensionality is particularly important in our experiment where current is injected at one wall only: because of this 
particularity, asymmetric three-dimensionality gives a measure of the influence of the forcing on the flow dynamics.\\ 
Our analysis will be restricted to the turbulent regimes, where the flow and 
the signals are more intense and the corresponding measurements more precise. 
{Since we consider variations of voltage with respect to the 
centrepoint of the vertical profile, the loss in precision on $\nabla\phi$ 
incurred by the lower flow intensity is compensated by the larger distance 
between electrodes, which is proportional to the amplitude of the signal (4 mm 
to 46 mm, \emph{vs.} 2.5 mm for gradients of potentials measured in the central 
square of the Hartmann walls). For sufficiently turbulent flows, the precision 
on $\nabla\phi$ therefore remains similar to that of measurements along Hartmann walls.} Profiles are chosen at 
$(x/h,y/h)={(-0.5,0.24)}$ and for $\lambda_i=0.1$ only, but are representative of the phenomenology observed in other cases.\\

\subsection{Symmetric \textit{vs.} antisymmetric three-dimensionality: influence of the forcing \label{sec:sym3d}}
\begin{figure}
\centering
\includegraphics[width=0.425\textwidth]{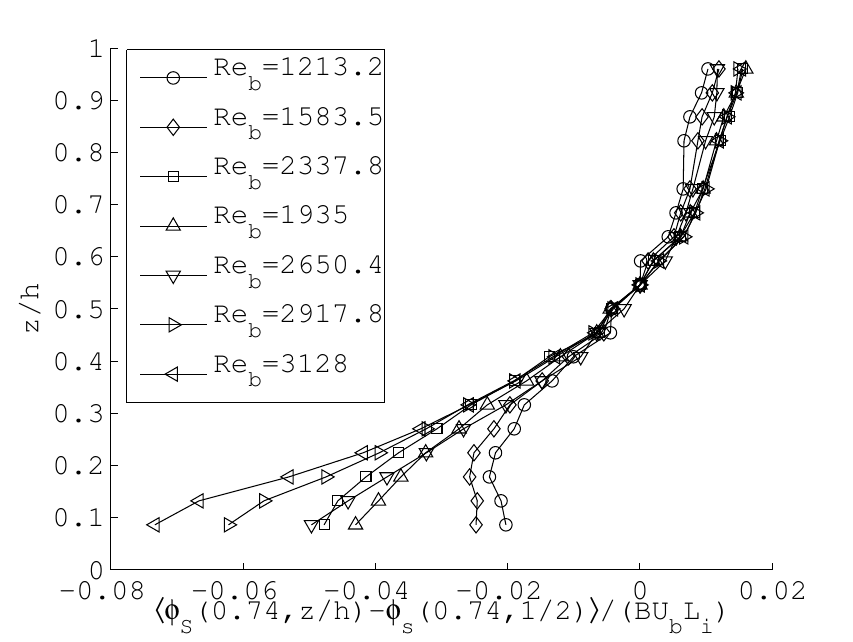}
\includegraphics[width=0.425\textwidth]{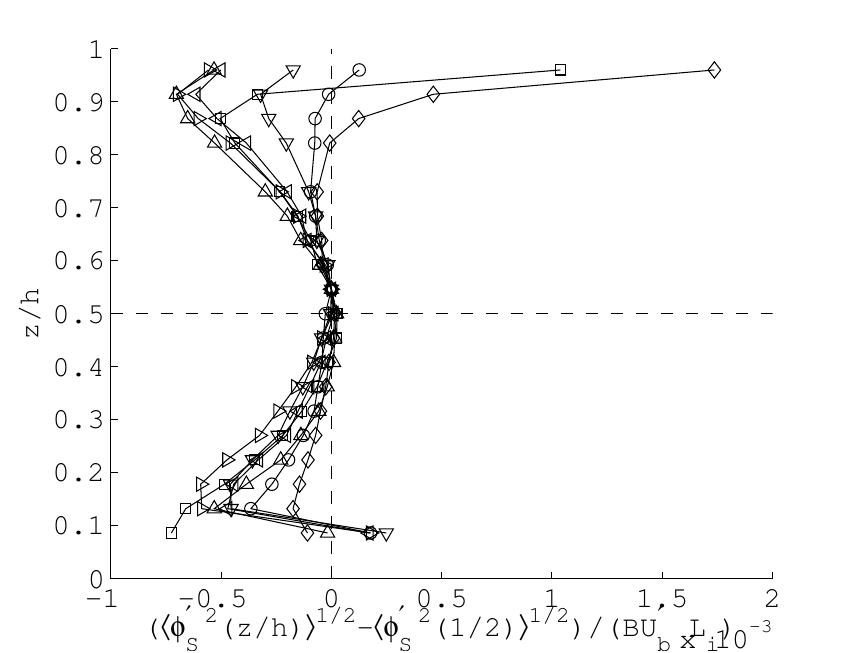}
\includegraphics[width=0.425\textwidth]{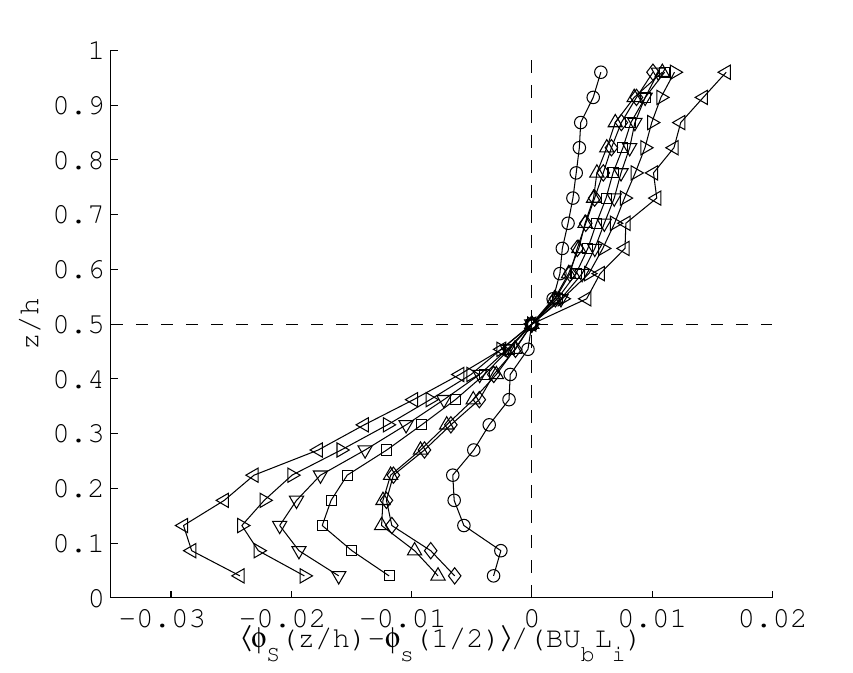}
\includegraphics[width=0.425\textwidth]{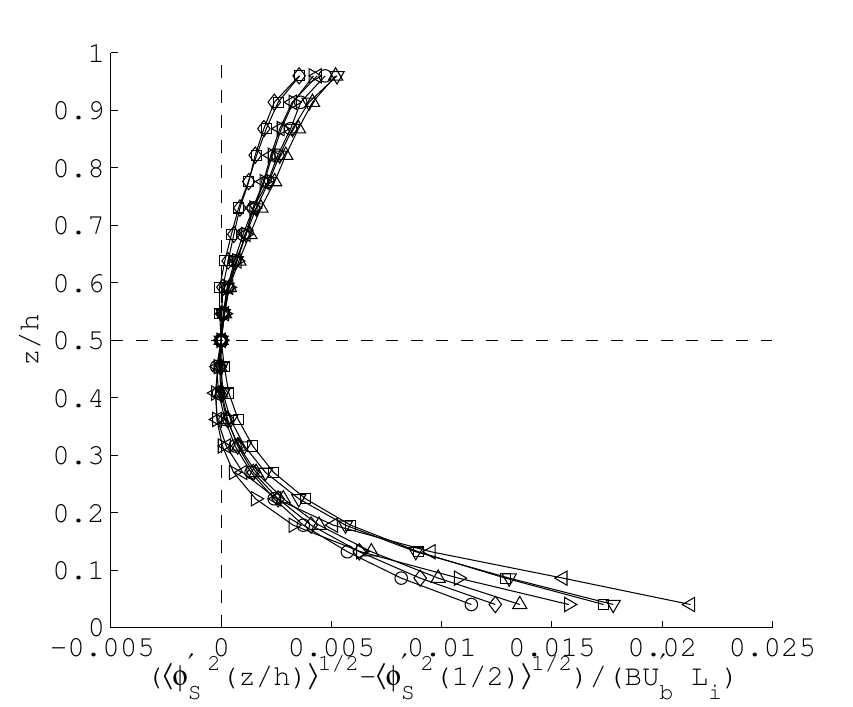}
\includegraphics[width=0.425\textwidth]{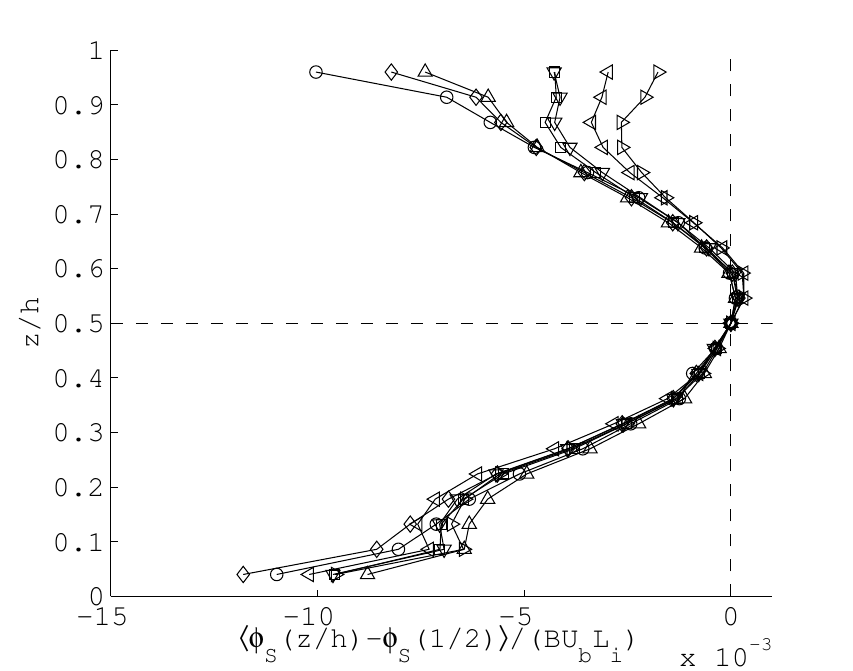}
\includegraphics[width=0.425\textwidth]{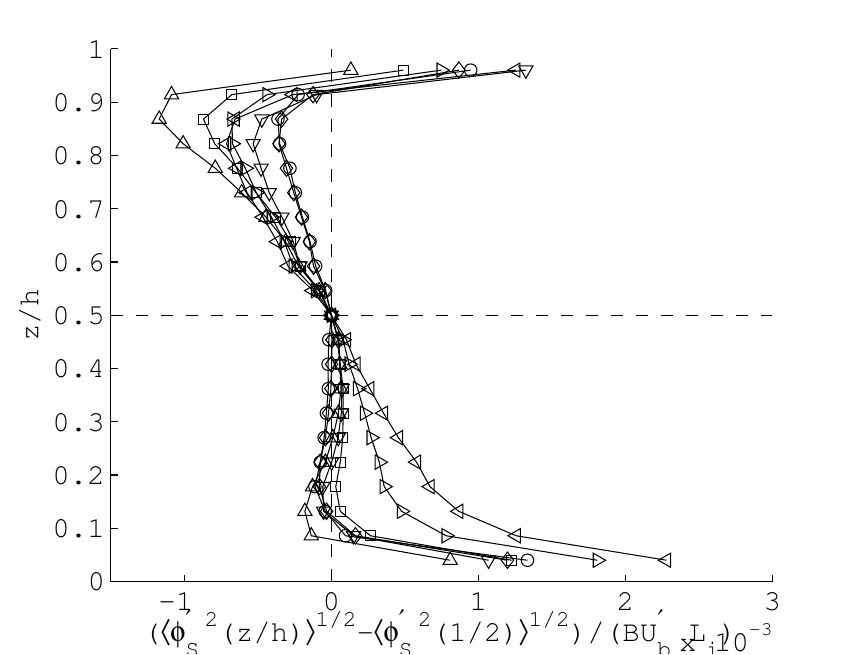}
\includegraphics[width=0.425\textwidth]{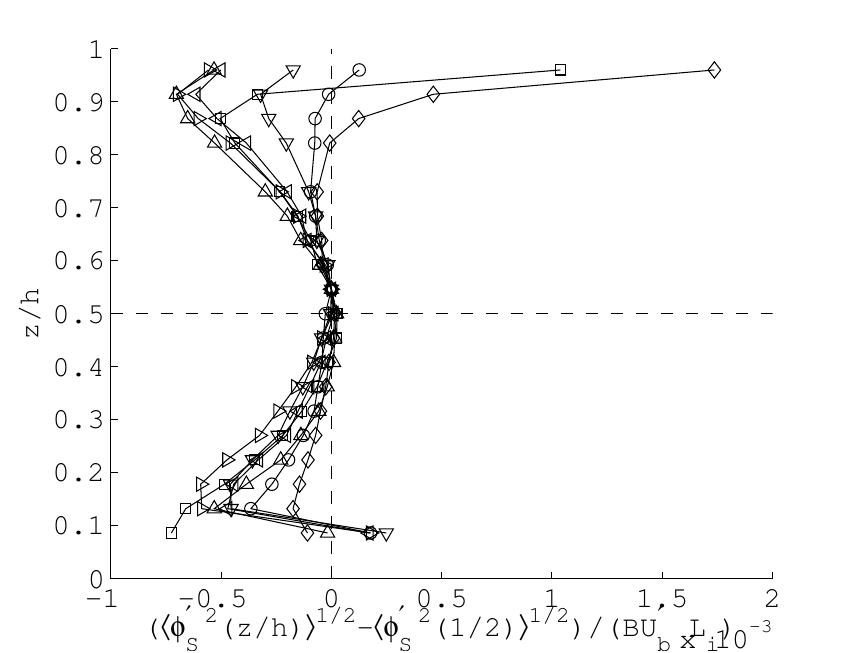}
\includegraphics[width=0.425\textwidth]{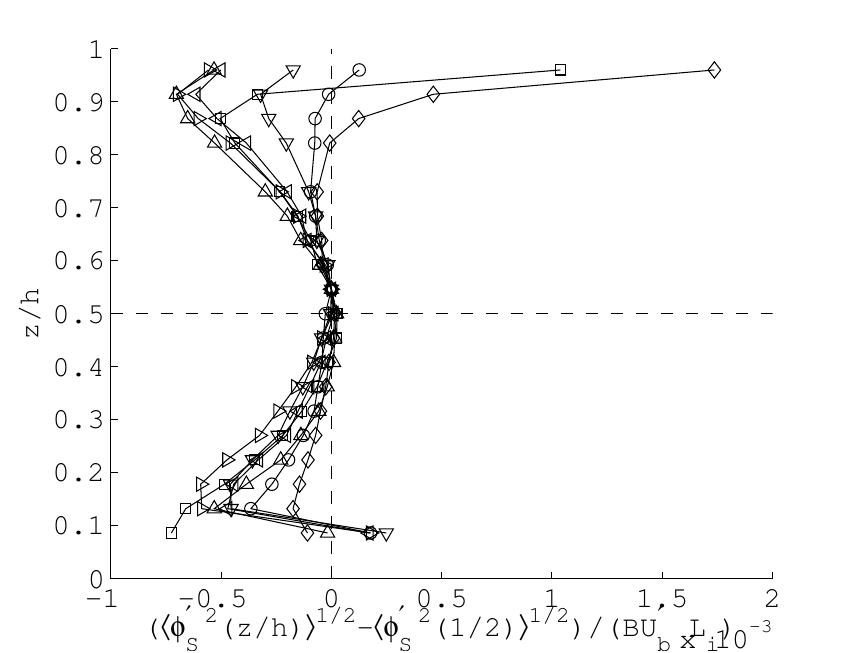}
\caption{Profiles of $\langle\phi_S(z/h)-\phi_S(z/h=1/2)\rangle$ (left), and 
$\langle\phi_S^\prime(z/h)^2\rangle^{1/2}-\langle\phi_S^\prime(z/h=1/2)^2\rangle^{1/2}$ (right) 
measured at $(x,y)/L=(-0.5,0.24)$. From top to bottom: $H\!a=$1822.2, 3644.3, 10933, 18222.}
\label{fig:phis}
\end{figure}

Figure \ref{fig:phis} shows the profiles of 
$\langle\phi_S(z/h)-\phi_S(1/2)\rangle$ and 
$\langle\phi_S(z/h)^{\prime2}\rangle^{1/2}-\langle\phi_S(1/2)^{\prime2}\rangle^{1/2}$ for the same set of values of $Re^0$, at different $H\!a$. They are 
respectively normalised by $B U_b L_i$ and $B U_b^\prime L_i$. First, 
 the amplitude of the variations of both these quantities 
decreases by about an order of magnitude between $H\!a=1822.2$ and $H\!a=18222$:
 the flow becomes closer to quasi-two dimensionality 
as $H\!a$ increases and the proportion of current that transits from the core to the Shercliff layers reduces correspondingly.
Second, all graphs show that 
the vertical gradient of these quantities increases with $Re^0$. This is due 
to the increasing amount of current drawn into the core by inertial effects, 
which loops back through the Shercliff layers. This effect being nonlinear, 
it can be measured on the graphs through the discrepancy between curves at different 
values of $Re^0$ for the same $H\!a$. At $H\!a=1822.2$, the eddy currents originating
 from the core are strongly concentrated near the bottom wall and practically 
zero  near the top wall (this is even more spectacular on the profiles of 
fluctuations). Together with the scaling $Re_t^\prime\sim(Re^0)^{1/2}$ observed 
in this regime  (section \ref{sec:ut_scalings}), 
this concurs to show that the top wall is not active and that $l_z^{\prime(N)}<h$. 
The phenomenology is similar at $H\!a=3644.3$, except for a slight gradient in 
the vicinity of the top wall, suggesting that $l_z^{\prime(N)}$ is closer to $h$ and 
that eddy currents already appear between the core and the top Hartmann layer.
At the two highest values of $H\!a$, the profiles of both averages and fluctuations become more symmetric, to be almost symmetric at $H\!a=18222$, but with an important difference between them. All average profiles converge towards a single 
curve at high $Re^0$: this indicates linearity with the flow intensity and 
therefore suggests that the corresponding three-dimensionality 
is due to the viscous effects in the Shercliff layers. By contrast, the 
profiles of fluctuations are more scattered: although the amplitude of the 
corresponding currents is quite weak, this nonlinearity suggests the presence 
of symmetric three-dimensionality in turbulent fluctuations, as theorised in section \ref{sec:s_as_3d}.\\
Symmetric three-dimensionality can be more precisely traced by inspecting the 
symmetric parts {$\phi_S^{(S)}$ and $\phi_S^{\prime (S)}$} of 
the profiles of $\langle\phi_S(z/h)-\phi_S(1/2)\rangle$ and
$\langle\phi_S(z/h)^{\prime2}\rangle^{1/2}-\langle\phi_S(1/2)^{\prime2}\rangle^{1/2}$, on figure \ref{fig:phis_sym}. While inertia-driven, symmetric 
three-dimensionality is practically absent from fluctuations at low $H\!a$, it 
becomes dominant a large $H\!a$. The average flow, by contrast, exhibits the 
opposite behaviour, with  inertia-driven symmetric three-dimensionality absent 
at high Hartmann number. Since the profiles of fluctuations 
are highly symmetric in this regime, this important difference shows 
that at $H\!a>10933$, where $\tau_{2D}$ becomes shorter than the turnover time 
 for the large scales $\tau_{u^\prime}$ {(or equivalently, when $N_t$ becomes greater than unity)}, the asymmetry induced by the forcing is mostly confined to 
the base flow and the Lorentz force damps it out before it is transferred to 
 turbulent fluctuations. The dimensionality of turbulent fluctuations generated 
in this regime is therefore subject to little influence from the inhomogeneity 
of the forcing.\\
\subsection{Experimental measure of the length of diffusion by the Lorentz force\label{sec:lz}}
\begin{figure}
\centering
\includegraphics[width=0.425\textwidth]{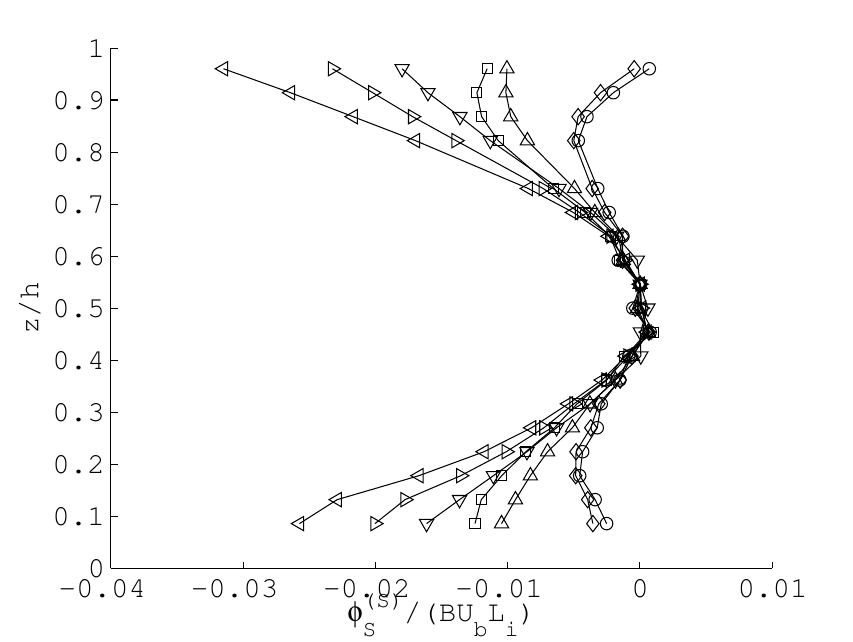}
\includegraphics[width=0.425\textwidth]{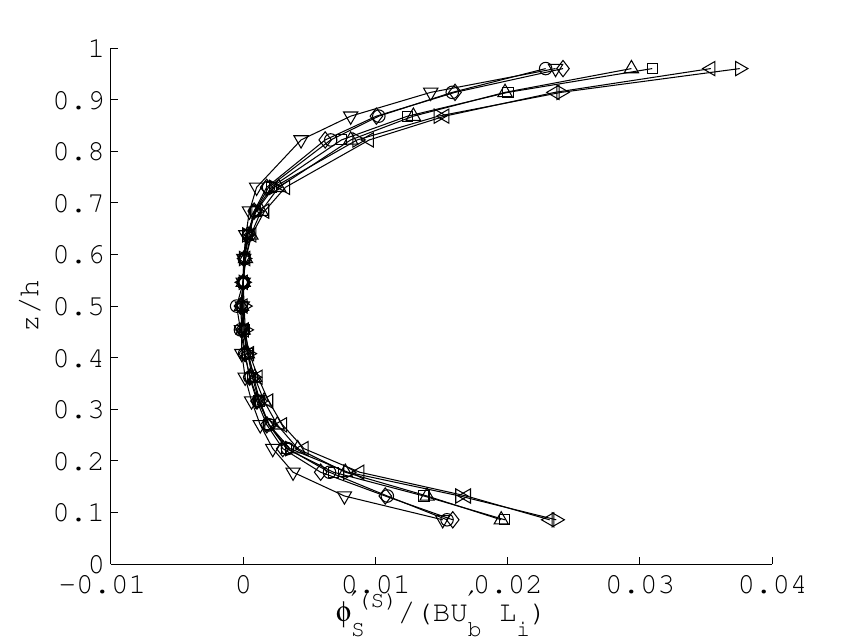}
\includegraphics[width=0.425\textwidth]{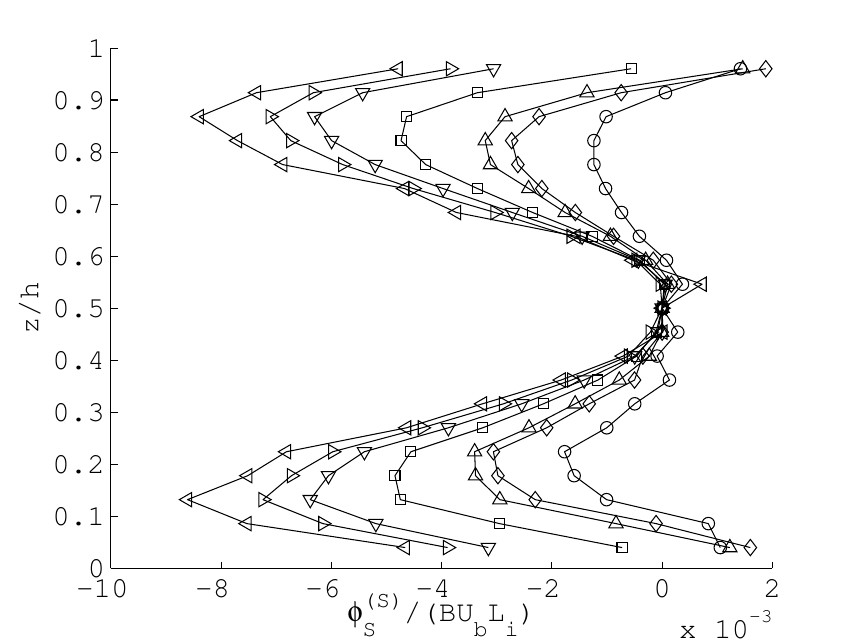}
\includegraphics[width=0.425\textwidth]{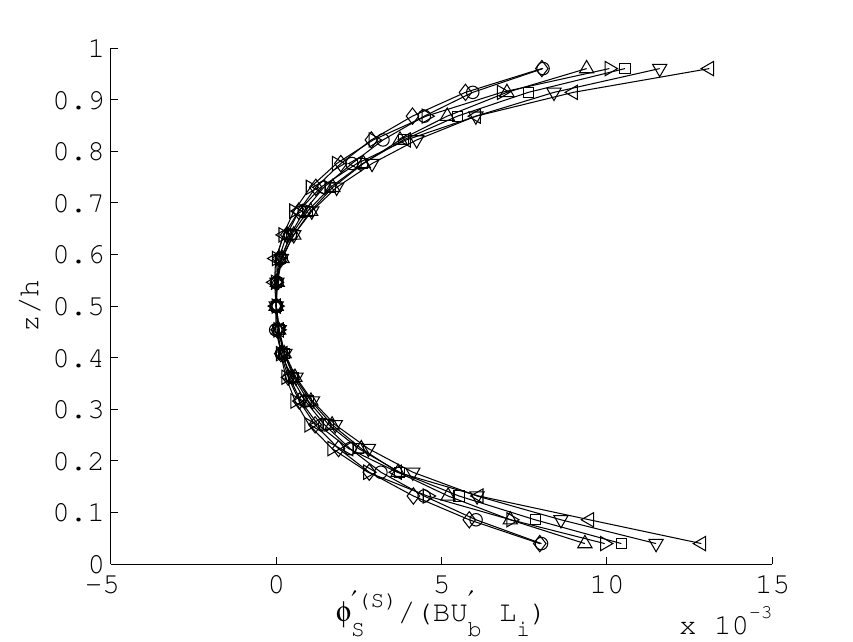}
\includegraphics[width=0.425\textwidth]{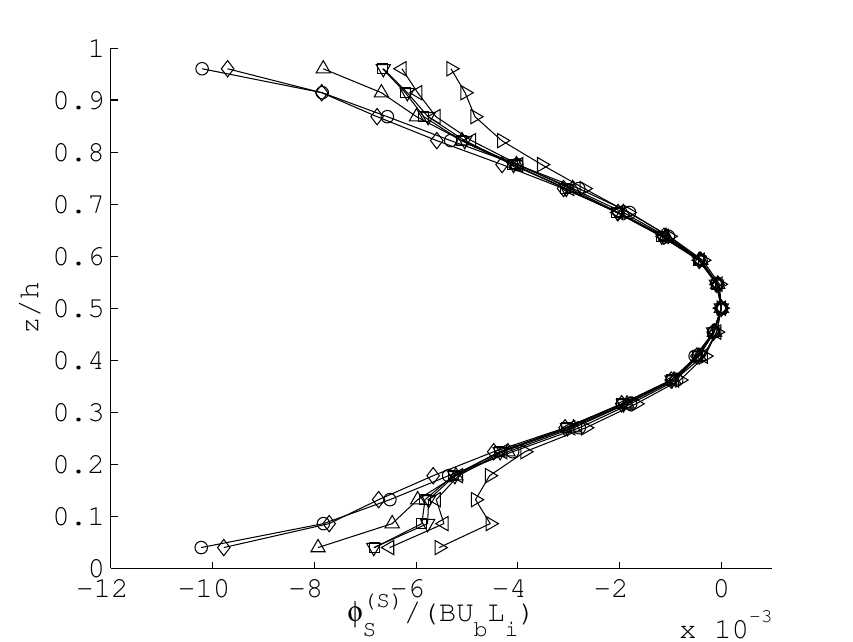}
\includegraphics[width=0.425\textwidth]{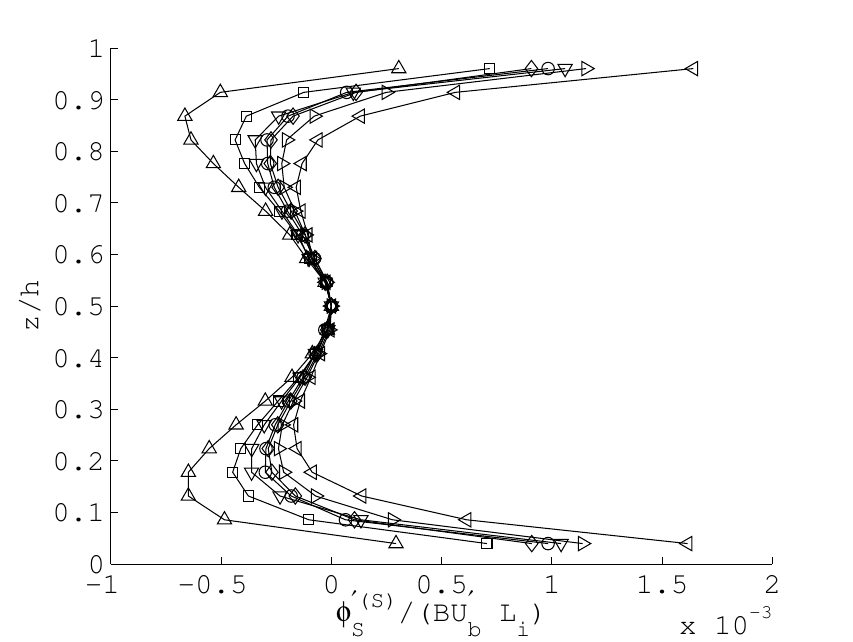}
\includegraphics[width=0.425\textwidth]{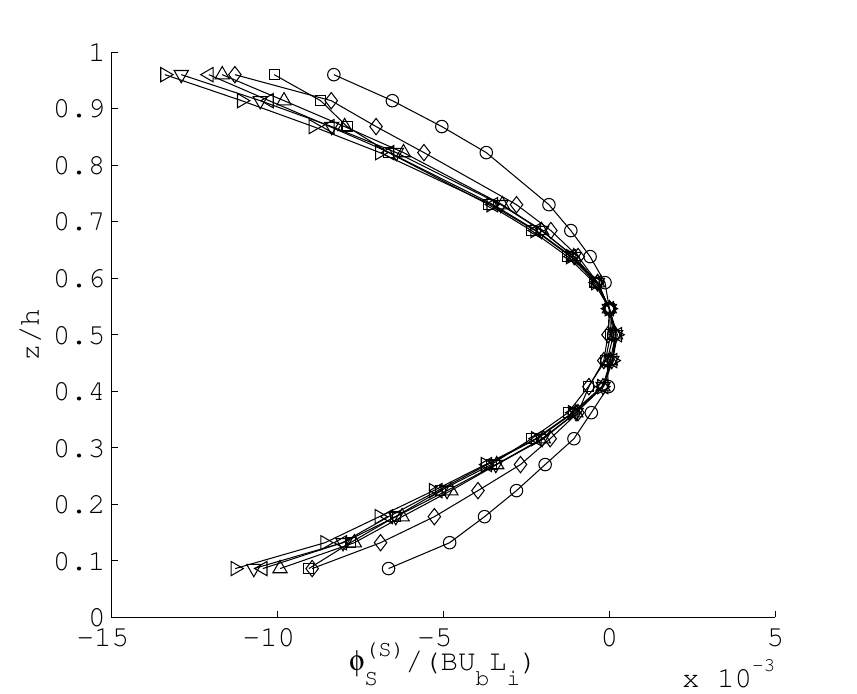}
\includegraphics[width=0.425\textwidth]{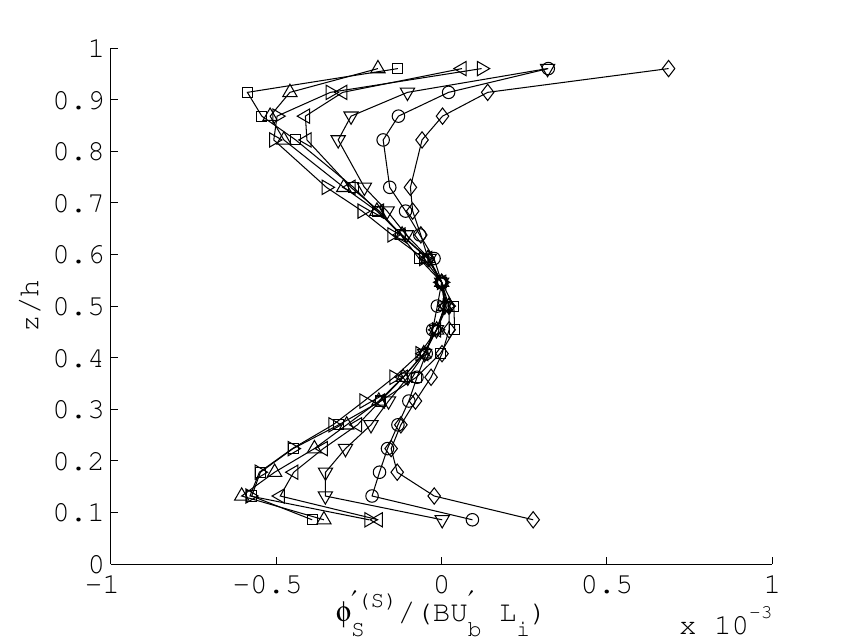}
\caption{Symmetric parts (superscript (S)) {$\phi_S^{(S)}$ and $\phi_S^{\prime(S)}$ respectively} of $\langle\phi_S(z/h)-\phi_S(z/h=1/2)\rangle$ (left), and of $\langle\phi_S^\prime(z/h)^2\rangle^{1/2}-\langle\phi_S^\prime(z/h=1/2)\rangle^{1/2}$ (right), measured at $(x/L,y/L)=(-0.5,0.24)$. From top to bottom: $H\!a=$1822.2, 3644.3, 10933, 18222, legend is in  figure \ref{fig:phis}.}
\label{fig:phis_sym}
\end{figure}

\begin{figure}
\centering
\includegraphics[width=0.8\textwidth]{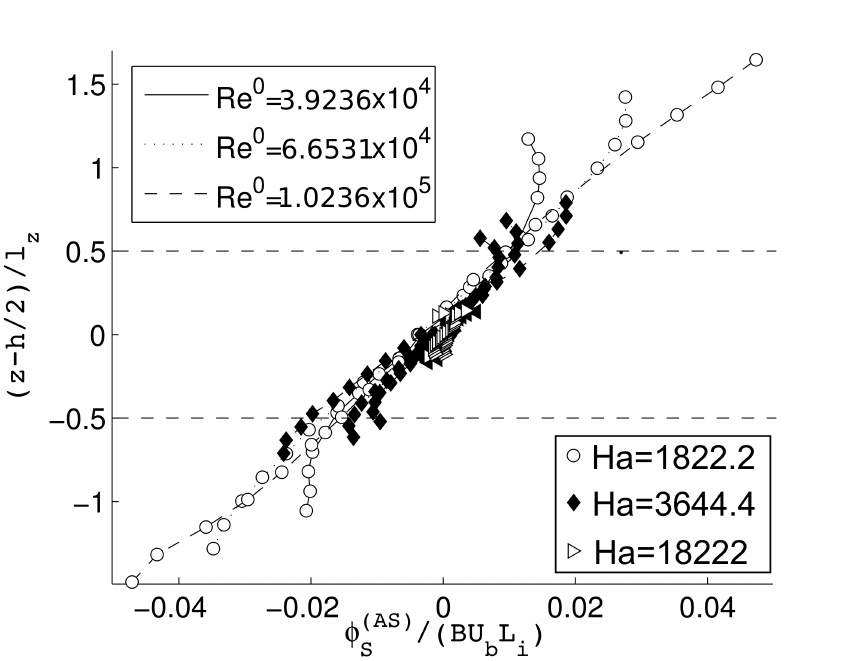}
\includegraphics[width=0.8\textwidth]{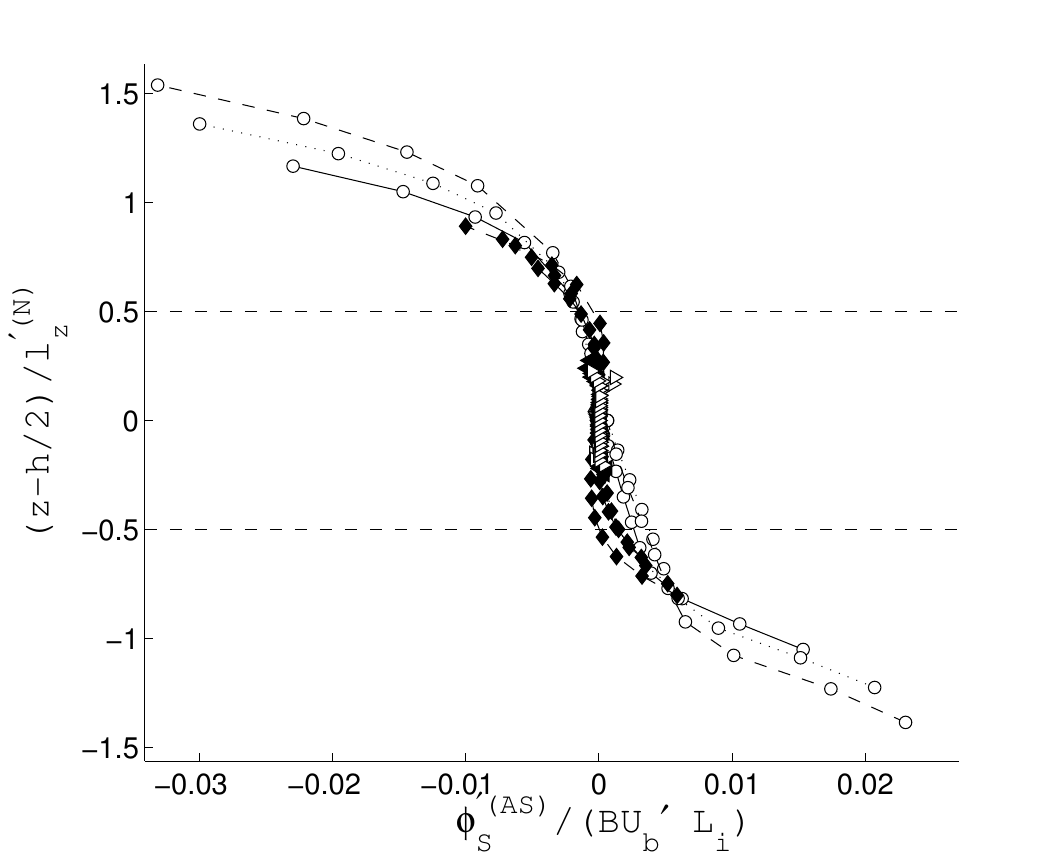}
\caption{Antisymmetric part (superscript (AS)) {$\phi_S^{(AS)}$ and $\phi_S^{\prime(AS)}$ respectively} of the profiles of $\langle\phi_S(z/h)-\phi_S(1/2)\rangle$ (top) and  $\langle\phi_S^\prime(z/h)^2\rangle^{1/2}-\langle\phi_S^\prime(z/h=1/2)^2\rangle^{1/2}$ (bottom), measured at $(x/L,y/L)=(-0.5,0.24)$, with $(z-h/2)$ respectively 
normalised by $l_z^{(N)}$ and $l_z^{\prime(N)}$  (built on $U_b$ and $U_b^\prime$). Profiles are plotted for $\lambda_i=0.1$, for 5 values of $H\!a$ 
and 3 values of $Re^0$ for each value of $H\!a$.}
\label{fig:phis_as_lzn}
\end{figure}

We shall now extract the lengthscale of momentum
diffusion by the Lorentz force $l_z^{(N)}(B,U_b)$, defined by 
(\ref{eq:lznonaxi}) from the antisymmetric parts 
{$\phi_S^{(AS)}$ and $\phi_S^{\prime (AS)}$} of the profiles of 
$\langle\phi_S(z/h)-\phi_S(1/2)\rangle$ and 
$\langle\phi_S(z/h)^{\prime2}\rangle^{1/2}-\langle\phi_S(1/2)^{\prime2}\rangle^{1/2}$. Profiles of these quantities are plotted on figure 
\ref{fig:phis_as_lzn} respectively against $(z-h/2)/l_z^{(N)}$ and $(z-h/2)/l_z^{\prime(N)}$. 
With this choice of variables, since $l_z^{(N)}$ and $l_z^{\prime(N)}$ increase with $H\!a$, 
curves obtained at higher $H\!a$ are shorter than those at low $H\!a$, and all 
curves are centred on 0.
The first important result is that profiles of average 
and fluctuations collapse well into a single curve each: this indicates that 
the momentum forced near the bottom wall reaches into the 
flow over $l_z^{(N)}$. This establishes the validity of scaling 
(\ref{eq:lznonaxi})  for $l_z^{(N)}$ and $l_z^{\prime(N)}$, for the average 
flow and the fluctuations respectively.\\
The average potentials follow a linear profile, with 
strong discrepancies to it in regions where $|(z-h/2)/l_z^{(N)}|>0.5$ 
(marked by horizontal dashed lines). Only at $H\!a=1822.2$ are these regions 
clearly reached, as $l_z^{(N)}$ is notably smaller than $h$ 
whereas at $H\!a=3644.3$, $l_z\simeq h$, depending on $Re^0$. Strong current 
exists in the corresponding regions of the core and so diffusion by the Lorentz 
force is not effective there. Since the inhomogeneity is asymmetric, it can be attributed to the forcing. In that sense, it 
carries the signature of the forcing dimensionality.
In the region $|(z-h/2)/l_z^{(N)}|<0.5$, by contrast, momentum diffusion by the Lorentz 
force dominates inertia: the (asymmetric) horizontally divergent current 
collapses in the core and therefore in the Shercliff layers, where it 
would normally return. 
For $H\!a>3644.3$, {$l_z^{(N)}> h$}: discrepancies to linearity are less 
pronounced and reflect eddy currents more homogeneously spread along the height 
of the vessel. This phenomenology holds for both the average flow and the 
fluctuations but the collapse to a single curve is significantly better on 
the profiles of fluctuations.  
It is remarkable that the antisymmetric part of turbulent fluctuations in the 
region $|(z-h/2)/l_z^{\prime(N)}|<0.5$ is dominated by momentum diffusion by 
the Lorentz force, when at the same time, three-dimensionality can still be observed in the symmetric part of the profiles.
Since for $H\!a>10933$, this region extends along the full height of the vessel, this implies that the three-dimensionality of turbulent fluctuations observed in this range of parameters (visible in particular on the symmetric profiles of figure \ref{fig:phis_sym}) is not induced by the forcing, 
even though the forcing still induces some three-dimensionality in the average flow. This is the second important result of this section. 
%

%
%

%

\section{Weak \emph{vs.} Strong three-dimensionality in unsteady regimes}
\label{sec:strong_weak}
\subsection{Global quantification of \emph{weak} and \emph{strong} three-dimensionality}
\begin{figure}
\centering
\psfrag{1mCp1}{$1-\overline{C^\prime_1}$}
\psfrag{1mCp2}{$1-\overline{C^\prime_2}$}
\psfrag{1mCp1law}{$1-C=0.5 N_t'^{-0.9}$}
\psfrag{1mCp2law}{$1-C=N_t'^{-0.9}$}
\psfrag{C1pr}{$\overline{C^\prime_1}$}
\psfrag{C2pr}{$\overline{C^\prime_2}$}
\psfrag{C1prlaw}{$C=0.5 N_t'^{1/2}$}
\psfrag{C2prlaw}{$C=0.25 N_t'$}
\psfrag{Nt}{$N_t'(L_i)$}
\psfrag{Co}{$C$}
\psfrag{1mCo}{$1-C$}
\includegraphics[width=8.5cm]{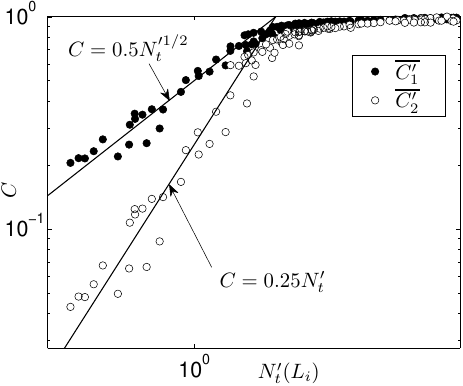}
\includegraphics[width=8.5cm]{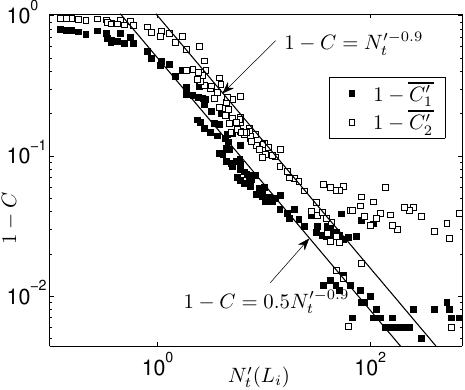}
\psfrag{a}{\begin{large} $N_t'=0.12$ \end{large}}
\psfrag{b}{\begin{large} $N_t'=0.42$ \end{large}}
\psfrag{c}{\begin{large} $N_t'=4.5$ \end{large}}
\psfrag{d}{\begin{large} $N_t'=65$ \end{large}}
\psfrag{bot}{\begin{large} bottom wall \end{large}}
\psfrag{top}{\begin{large} top wall \end{large}}
\includegraphics[width=9cm]{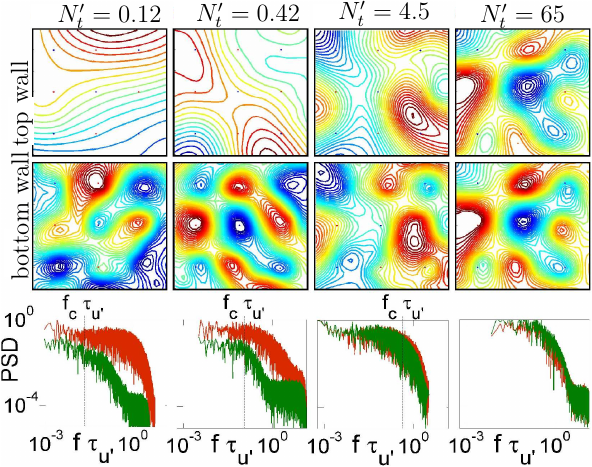}
\caption{Correlations (top) and co-correlations (middle) of electric potential gradient between bottom and top Hartmann walls, averaged over the central square. Data were obtained from measurements both with $\lambda_i=0.1$ and $\lambda_i=0.3$. Bottom: snapshots of contours of electric potential taken for four values of $N_t'$ and examples of spectra taken pairs of probes within the central square respectively from top (green) and bottom plate (red), at the same point $(x,y)$.
\label{fig:corr}}
\end{figure}

In sections \ref{sec:theory}, \ref{sec:visc-inertia}, and \ref{sec:wall_scalings}, the vertical profiles 
 of electric potential and the scalings for $U_b$, $U_t$, 
$U_b^\prime$ and $U_t^\prime$ were shown to be efficient tools to detect  
three-dimensionality, and to contain the viscous or inertial signature of its origin. 
We shall now more precisely measure three-dimensionality and determine its nature by quantifying weak and strong three-dimensionalities,
 which we previously introduced in \cite{kp10_prl}. 
 In particular, we shall estimate the extent to which either form of three-dimensionality is eradicated in the quasi-two dimensional limit.\\
Let us start by recalling that 
\emph{weak} three-dimensionality is characterised by a variation in flow 
intensity between two-dimensional "slices" of topologically equivalent flows in the $(x,y)$ plane. \emph{Strong} three-dimensionality, by contrast, involves 
different profiles of physical quantities along $z$ at different 
locations $(x,y)$. Following \cite{kp10_prl}, their respective occurrence shall 
be quantified by comparing measurements at bottom and top Hartmann walls 
at $z=0$ and $z=h$, by means of two types of correlations:
\begin{eqnarray}
C'_1(x,y)&=&\frac{\sum\limits_{t=0}^{T}\partial_y \phi'_{\rm b}(x,y,t) \partial_y \phi'_{\rm t}(x,y,t)}{\sqrt{\sum\limits_{t=0}^{T} \partial_y {\phi'}^2_{\rm b}(x,y,t) \sum\limits_{t=0}^{T}
\partial_y {\phi'}^2_{\rm t}(x,y,t)}}, \label{eq:c1p}\\ 
C'_2(x,y)&=&\frac{{\sum\limits_{t=0}^{T}} \partial_y \phi'_{\rm b}(x,y,t) \partial_y \phi'_{\rm t}(x,y,t)}{{\sum\limits_{t=0}^{T} \partial_y {\phi'}^2_{\rm b}(x,y,t) }},
\label{eq:c2p}
\end{eqnarray}
where $T$ is the duration of the recorded signals. Values of $C'_1$ depart from 
unity whenever strong three-dimensionality is present, but are unaffected by 
weak three-dimensionality. $C'_2$, on the other hand, differs from unity 
whenever either weak or strong three-dimensionality is present. In order to 
minimise the influence of the side walls, we shall reason on the 
spatial averages $\overline{C'_i}$ and $1-\overline{C'_i}$ of $C'_i$ and 
$1-C'_i$ over the central square region (see section \ref{sec:setup}).
Whilst $1-C'_1$ and $1-C'_2$ give a measure of three-dimensionalities, 
$C'_1$ and $C'_2$ reflect the emergence of two-dimensionality, respectively in 
the strong sense, and in both the weak and the strong sense. A minor limitation 
of this method, however, is that fluctuations that remain \emph{at all time} 
symmetric with respect to the vessel mid-plane ($z/h=1/2$) cannot be detected 
as they do not affect the values of $C_1^\prime$ nor $C_2^\prime$. Fluctuations 
that are symmetric \emph{on average} but not at all time, such as those 
detected in the symmetric part of $\phi^\prime_S$ (section 
\ref{sec:wall_scalings}), may, by contrast alter the values of the 
correlations. While the former type 
of fluctuation is unlikely to generate strong three-dimensionality, the latter 
is the most likely form of \emph{strong} symmetric three-dimensionality, since
it results from randomly localised three-dimensional disturbances, which are 
unaffected by the asymmetry of the forcing and therefore distributed 
symmetrically along the vessel height. {The experimental precision on $\overline{C'_i}$ can be estimated from the ratio of signal amplitude to 
uncorrelated noise. Using definitions (\ref{eq:c1p}) and (\ref{eq:c2p}), we find that 
the relative theoretical precision is never worse than 2\%. It concerns regimes of low $H\!a$ were the signal is lower and 
the values of $\overline{C'_i}$ close to zero. For the highest values of $H\!a$, where 
$\overline{C'_i}$ becomes very close to 1, the uncertainty falls around that 
incurred by the mechanical tolerances of the vessel (about 0.1\%). In practise, we found it to be around 0.5\%. }
\\ 
$\overline{C'_i}$ and  $1-\overline{C'_i}$ are plotted in figure \ref{fig:corr} 
 against the true interaction parameter based on the velocity fluctuations and 
the injection scale 
$N_t'(L_i)=(\sigma B^2 L_i/\rho U^\prime_b)\lambda_i^{-2}$. All our 
measurement points collapse into two single
curves $\overline{C'_1}({N}_t)$ and $\overline{C'_2}({N}_t)$, indicating  
that three-dimensionality in the fluctuations is exclusively of inertial 
nature. This is consistent with the scalings found for 
fluctuations $U_b^\prime$ and $U_t^\prime$ in section \ref{sec:visc-inertia}, 
which reflected the presence of inertia in the core too. 
In the limit $N_t'\rightarrow\infty$, both types of correlations become 
close to 1 {within} a precision of about 1\%, indicating that in these regimes, 
the flow is very close to quasi-two-dimensionality. Furthermore, 
in the limit $N_t'\rightarrow\infty$, $1-\overline{C'_1}({ N}_t)\sim N_t'^{-0.9}$,
 and $1-\overline{C'_2}({ N}_t)\sim N_t'^{-0.9}$, up to the point where  
these quantities reach a plateau reflecting the level of signal noise. In fact two such plateaux shoot off from the main $N_t'^{-0.9}$ law, that 
correspond to the two different levels of signal to noise ratio for 
$\lambda_i=0.1$ and $\lambda_i=0.3$. {Despite their purely empirical nature, these scalings give} a strong experimental evidence that the flow 
becomes \emph{asymptotically} quasi-two dimensional and that a vanishingly 
small amount of three-dimensionality always remains present. 
By contrast, the transition between two and three dimensional states in 
MHD flows with non-dissipative boundaries (such as free-slip conditions studied 
by \cite{thess07_jfm} or periodic boundary conditions by 
\cite{zikanov98_jfm,pdy10_jfm}) occurred through bifurcations at a critical 
value of the parameter representing the flow intensity. This fundamental 
difference 
in behaviour must be attributed to the role of walls, in line with our 
recent theory (\cite{p12_epl}) showing that walls 
incur weak three-dimensionality as soon as $N$ (or $N_t$) is finite.\\
In the limit $N_t'\rightarrow0$, $\overline{C'_1}$ and $\overline{C'_2}$ exhibit several noticeable 
differences. Firstly, {measurements empirically obey two 
different scalings: $\overline{C'_1}\sim N_t'^{1/2}$, and $\overline{C'_2}\sim N_t'$.
This indicates that weak three-dimensionality is more significant than 
 strong three-dimensionality}. This 
effect is in part due to the nature of the 
forcing: for low magnetic fields and strong currents, electric currents remain 
in the vicinity of the bottom wall, where the flow is intensely stirred. By 
contrast, very little current flows near the top wall.  Whatever  
weak flow remains there is mostly entrained by viscous momentum diffusion from 
the intense flow near the bottom wall. 
Both share the same large patterns, as on the snapshots of streamlines
 (a) and (b) from figure \ref{fig:corr}, and are therefore somewhat 
correlated in the weak sense. By contrast, the very small intensity of the 
flow near the top wall extinguishes correlation in the strong sense.\\
The power laws found in the limits $N_t'\rightarrow0$ and $N_t'\rightarrow\infty$
intersect at $N_t'=N_t'^{2D}\simeq4.5$, and remain remarkably valid in their 
respective ranges $N_t'<4.5$ and $N_t'>4.5$, 
In this sense, the value $N_t'^{2D}$ 
marks a form of transition between quasi-two dimensional and three dimensional 
states. We insist, 
however, that this isn't a critical value of $N_t'$ since no bifurcation occurs 
in the flow states at this point. Also, it provides a rather global 
appraisal of the flow state but it is certainly not the case that 
for $N_t'>4.5$, every structure in the flow is quasi-two dimensional.
\subsection{Frequency dependence of strong three-dimensionality}
\begin{figure}
\centering
\psfrag{ff}{$f$ (Hz)}
\psfrag{fc}{$f_c$}
\psfrag{cp1f}{$c^\prime_1(f)$}
\psfrag{ftup}{$f\tau_{u^\prime}$}
\psfrag{fctup}{$f_c\tau_{u^\prime}$}
\psfrag{dphit}{\tiny $\partial_y\phi^\prime_t$}
\psfrag{dphib}{\tiny $\partial_y\phi^\prime_b$}
\psfrag{bdphit}{\tiny $\mathcal B_f\partial_y\phi^\prime_t$}
\psfrag{bdphib}{\tiny $\mathcal B_f\partial_y\phi^\prime_b$}
\psfrag{fleqfc}{\small $f=0.05 {\rm Hz } <f_c$}
\psfrag{feqfc}{\small $f=0.14 {\rm Hz } =f_c$}
\psfrag{fgtfc}{\small $f=0.5 {\rm Hz } >f_c$}
\includegraphics[width=13.5cm]{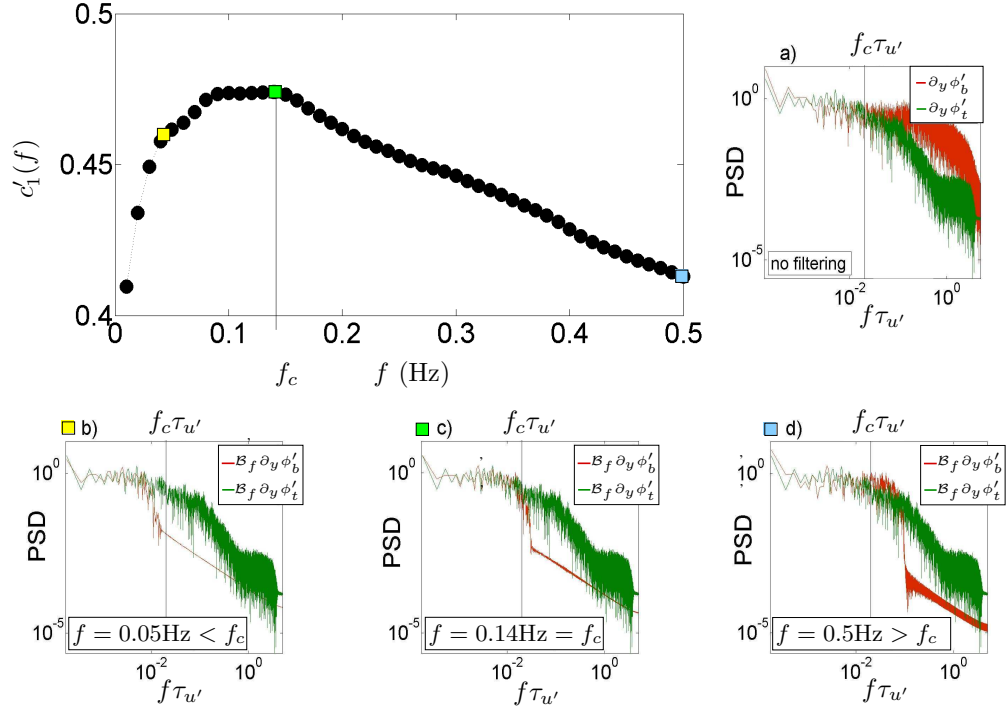}
\caption{Variations of partial correlation $c'_1(f)$ with the cutoff frequency of filter $\mathcal B_f$ applied to the signal measured at the bottom 
wall. Curves are obtained for $\lambda_i=0.1$ and $N_t'=0.12$ for a pair of 
probes within the central square. Curves obtained 
for different parameters exhibit the same features. {a)-d): 
spectra of the filtered signals and top signals plotted in terms of the 
frequency normalised by the turnover frequency associated to RMS fluctuations $\tau_{u^\prime}^{-1}$. Frequency $f_c$ is also represented}.}
\label{fig:partial_corr}
\end{figure}

\begin{figure}
\centering
\psfrag{L1}{$\lambda_i=0.1$}
\psfrag{L3}{$\lambda_i=0.3$}
\psfrag{Nt}{$N_t'$}
\psfrag{F}{$f_c\tau_{u^\prime}N_t'^{-2/3}$}

\centering
\includegraphics[width=12cm]{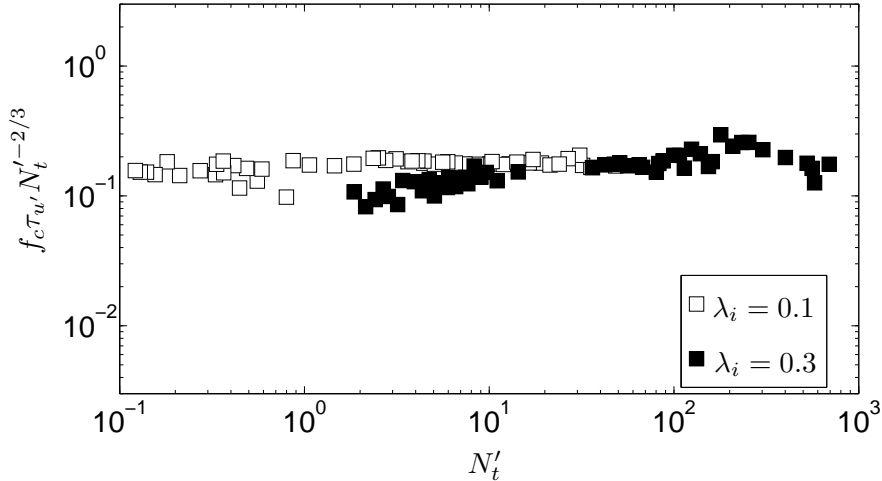}
\caption{Cut-off frequency $f_c$, separating quasi two-dimensional fluctuations from 
three-dimensional ones,  normalised by $\tau_{u'}^{-1}$, the eddy turnover frequency built on $U_b^\prime$, \textit{vs.} $N_t'$.}
\label{fig:cutoff}
\end{figure}

We shall now seek to characterise three-dimensionality locally  
 by analysing the frequency spectra from the fluctuations of 
$\nabla\phi$ near each of the Hartmann walls. In figure \ref{fig:corr}, pairs of 
such spectra are shown that were obtained from signals acquired at locations 
of either Hartmann walls exactly aligned with the magnetic field lines 
({\textit i.e.} respectively at $z=0$ and $z=h$, but at the same coordinates 
$(x,y)$). These were taken in the central square region, so as to minimise 
the influence of the side walls. {At a sampling frequency of 
128 Hz, signals would theoretically render fluctuations of up 64 Hz. To 
increase signal quality, however, we apply a low-pass filter at 45 Hz so as to avoid electromagnetic interferences from the mains. The physically relevant part of the spectra correspondingly extends approximately over 2-3 decades.}
For values of $N_t'$ noticeably higher than 4.5, top and bottom spectra are practically 
identical, as would be expected of a flow that would be quasi-two dimensional 
at all scales. Since spectra do not carry any phase information, however, 
identical spectra do not provide a proof that all scales are indeed quasi 
two-dimensional, but merely that no obvious three-dimensionality is present at 
any scale.
Pairs of spectra at lower values of $N_t'$, on the other hand, show a clear 
manifestation of three-dimensionality as fluctuations of higher frequency carry 
significantly less energy in the vicinity of the top wall than near the bottom 
wall. Fluctuations of lower frequency, by contrast, carry 
the same amount of energy. When $N_t'$ is reduced, the range of frequencies 
affected by three-dimensionality widens to the low-frequency range and ends 
up contaminating the whole spectrum at the lowest values of $N_t'$. 
This qualitative comparison of spectra indicates that fluctuations of higher 
frequencies, and smaller scales by extension,  are more sensitive to 
three-dimensionality than those of lower frequencies and larger scales, 
in accordance with \cite{sm82}'s prediction. A closer inspection of 
superimposed spectra further suggests the existence of a cutoff frequency $f_c$ 
separating three-dimensional high frequency fluctuations from lower 
frequency fluctuations with the same amount of energy near both walls. It 
also appears that $f_c$ increases with $N_t'$.
To quantify the behaviour of $f_c(N_t')$, 
we define a partial correlation function between signals on the bottom and top Hartmann walls in the strong sense: 
\begin{eqnarray}
c'_1(f)&=&\frac{\sum\limits_{t=0}^{T}\mathcal B_f\partial_y \phi'_{\rm b}(t) \partial_y \phi'_{\rm t}(t)}{\sqrt{\sum\limits_{t=0}^{T}\mathcal B_f \partial_y {\phi'}^2_{\rm b}(t) \sum\limits_{t=0}^{T}
\partial_y {\phi'}^2_{\rm t}(t)}}.
\end{eqnarray}
The definition of $c'_1(f)$ 
hardly differs from that of $C'_1(f)$
, but for the fact that the full signal recorded from the bottom 
wall is replaced with a filtered counterpart, processed through an $8^{th}$ 
order low-pass filter of cutoff frequency $f$, incurring neither phase nor 
amplitude distortion {(denoted $\mathcal B_f$)}. 
Figure \ref{fig:partial_corr} shows the typical  variations of $c'_1(f)$ 
when $f$ spans the whole part of the spectrum resolved in our measurements. 
Their most important feature is the presence of a maximum, which
can be understood by considering two spectra obtained from top and bottom walls  overlapping up to a frequency $f_c$. Increasing $f$ from 0 for $f<f_c$ adds fluctuations to the bottom signal that are correlated with existing frequencies in the signal from the top wall, so $c'_1(f)$ increases. When $f$ is increased 
beyond $f_c$, the fluctuations added to the bottom signal are decorrelated 
from existing frequencies in the top one, so $c'_1(f)$ decreases. 
The position of the maximum thus gives a good measure of $f_c$. 
Note that the same approach cannot be applied to $C_2^\prime$, as a partial 
correlation $c_2(f)^\prime$ obtained from $C_2^\prime$ by replacing the bottom 
signal by a filtered counterpart would diverge in the limit $f\rightarrow0$. 
This is because the filtered bottom signal vanishes in this limit (This 
feature is absent from the behaviour of $c_1(f)^\prime$ as signals are 
normalised by their intensity by construction). 
Nevertheless, from \cite{sm82}'s theory, 
the existence of a cutoff lengthscale {$k_c$}
separating quasi-two dimensional from three-dimensional structures stems from 
no assumption on the nature of inertial effects so it can be expected to exist 
both for weak and strong three-dimensionality. Yet, if instabilities 
occurred at the scale of individual structures, they would incur strong 
rather than weak three-dimensionality so in any case, $f_c$ is linked to the 
minimum size of structures that are stable to three-dimensional perturbations.
The question of the nature of the possible instabilities that occur below this 
size remains to be clarified with dedicated analysis.\\
The variations of $f_c(N_t')$ are represented in figure \ref{fig:cutoff}. Measurements points across the whole range of unsteady flows reachable 
in the experiment once again collapse into a single curve, so $f_c$ is 
determined by $N_t'$, as $\overline{C'_1(f)}$ and $\overline{C'_2(f)}$ were, 
with a scaling law
\begin{equation}
f_c\backsimeq 1.7 \tau_{u^\prime}^{-1} { N}_t ^{0.67},
\label{eq:fc}
\end{equation}
where $\tau_{u^\prime}=L_i/U_b^\prime$ is the turnover time at the forcing scale.
While this law is verified to a great precision for $\lambda_i=0.1$ (the exponent 0.67 is obtained to a precision of $\pm 0.03$), a small departure
 to it is visible at low $N_t'$ for $\lambda_i=0.3$, that must be attributed to the 
greater influence of the walls. Indeed the large vortices forced 
in such regimes can be expected to induce local separations of the parallel 
boundary layers, acting as a source of instability that could lower $f_c$. 
Unfortunately, our measurements do not allow us to test this hypothesis further. It is however consistent with the fact that boundary-layer separation is suppressed at higher $N_t'$ (\cite{psm05_jfm}). {The fact that this 
empirical scaling is reasonably independent of the forcing scale gives good 
evidence of its universality. This important result brings a phenomenological 
confirmation of \cite{sm82}'s theory, even though (\ref{eq:fc}) is not strictly equivalent to these authors' scaling for a cutoff wavelength $k_c$, of the form $k_c\sim N_t'^{1/3}$.}
 
%
\section{Conclusions}
We have conducted a detailed analysis of the conditions of appearance of 
three-dimensionality in low-$R\!m$ MHD turbulence in a channel bounded by two 
solid walls. We have clarified the mechanisms governing this phenomenon and were able to quantify it through the scaling of flow quantities. These results can be summarised as follows.
\subsection{The force driving three-dimensionality determines the scalings for the Reynolds number near Hartmann walls}
Three-dimensionality 
appears whenever viscous or inertial forces exist in the core of the flow, 
\emph{i.e.} outside the Hartmann boundary layers. The nature of these forces 
was identified through scaling laws linking the total current injected to 
drive the flow and the velocities in the vicinity 
of the bottom wall (where the flow is forced by injecting electric current), 
and the top wall. Driving the flow in this way offers a convenient way of 
controlling the external force generating the flow. Scaling laws expressing the response of the flow to the forcing were written in terms of Reynolds numbers $Re^b$, $Re^t$, against the forcing intensity measured by $Re^0$.
These yield two conclusions regarding both the average flow and for turbulent fluctuations:
\begin{itemize}
\item Both quasi-two-dimensional flows and flows where three-dimensionality is 
driven by viscous friction in the core obey scaling laws of the form first 
put forward by \cite{sommeria88}: $Re^b\sim Re^0$. We gathered both in a regime which we called \emph{inertialess}.
\item For flows where three-dimensionality is purely driven by inertia, in 
particular turbulent ones, $Re^b\sim (Re^0)^{2/3}$.
\end{itemize}
\subsection{In a channel of height $h$, three-dimensionality is determined by the ratio of the length of diffusion of momentum by the Lorentz force $l_z$ to $h$}
Momentum diffusion along $\mathbf B$ is opposed by either viscous or inertial 
forces and 
this determines the thickness $l_z$ of the fluid layer where momentum diffuses. If viscous effects oppose the Lorentz force, then $l_z\sim l_\perp H\!a$, whereas when inertia opposes it, $l_z\sim l_\perp N^{1/2}$. For a channel of height $h$, two regimes can be distinguished:
\begin{itemize}
\item If $h<l_z$, eddy currents flow in the top Hartmann layer and the upper wall is active in the sense that it strongly influences the flow. $Re^t$ then scales as $Re^t\sim Re^0(1-D^{(H\!a)}\lambda_i^2H\!a^{-1})$ in the inertialess regime, and as $Re^t\sim (Re^0)^{2/3}(1-D^{(N)}N_t^{-1/2})$ in the inertial regime. The flow becomes quasi-two-dimensional in the limit $l_z/h\rightarrow\infty$ (When the top wall is active, but the flow is still three-dimensional, inertia also incurs a small correction to the law $Re^b\sim (Re^0)^{2/3}$ that further reduces $Re^b$ slightly.).
\item  If $h>l_z$, momentum diffusion doesn't reach the top wall and only 
a weak residual flow exists there. In this sense the top wall is passive and 
the flow there satisfies the scaling found by \cite{duran-matute10_pre} of 
$Re^t\sim (Re^0)^{1/2}$.
\end{itemize}
These scalings were verified experimentally for two injection scales ($\lambda_i=0.1$ and 0.3),  $1093.3\leq H\!a \leq 18222$ and $Re^0$ up to $1.03\times10^5$, which corresponded to a Reynolds number based on turbulent fluctuations $Re^{b\prime}$ of up to about $6\times10^3$. 
{Exponents in the scalings for Reynolds number near Hartmann 
walls are recovered both experimentally and from a generic theory so they may 
be deemed a universal feature of MHD turbulence in channels. 
The same may not hold true for the multiplicative constants, which can be expected to depend on the geometry of the forcing.}\\

The average flow was found to be in the inertialess regime exclusively when 
the flow was steady and to exhibit inertia-driven three-dimensionality in the limit of $Re^0\rightarrow\infty$.
The fluctuating part of the unsteady flow, by contrast, exhibits a transition 
between the inertialess regime and the regime of inertial three-dimensionality. This difference between average and fluctuations singles out a range of forcing intensities where inertial three-dimensionality due to the forcing affects only the average flow but not turbulent fluctuations.\\
%
\subsection{Three-dimensionality was observed independently of the dimensionality of the forcing}
We were then able to track the three-dimensionality due to the electric 
forcing by separating symmetric and antisymmetric parts of profiles of electric 
potential measured along Shercliff walls. This led us to identify two basic features of low $R\!m$ MHD turbulence:
\begin{itemize}
\item Diffusion of momentum by the Lorentz force is indeed effective over the length theorised by \cite{sm82}, which we showed to be rather precisely $l_z \simeq N^{1/2}$ both for the average flow and for turbulent fluctuations.
\item When $\tau_{2D}(\lambda_i)\lesssim\tau_{U}(\lambda_i$), 
the asymmetry between 
the top and bottom wall introduced by the forcing was dampened out during 
the energy transfer from the mean flow to the large turbulent scales. In this 
case, the trace of the forcing was mostly borne by the average flow. The 
features of the turbulence observed in the flow fluctuations are then 
reasonably independent of it. 
In this sense, the three-dimensionality observed in this regime is intrinsic and not induced by the forcing.
\end{itemize}
\subsection{Three-dimensionality vanishes asymptotically in the quasi-two-dimensional limit}
Finally, in the turbulent regimes, we were able to quantify \emph{weak} and \emph{strong} 
three-dimensionalities, which we first introduced in \cite{kp10_prl}, both 
globally and in a frequency analysis of the 
flow. This brought further clarification on the mechanisms driving 
three-dimensionality. 
\begin{itemize}
\item The transition between the quasi-two-dimensional and three-dimensional 
states of wall-bounded MHD turbulence is a progressive phenomenon controlled by 
the true interaction parameter built on the large scales $N_t'$. It doesn't 
occur at a bifurcation in the space of parameters, unlike in domains with 
slip-free boundaries only. Instabilities of individual structures are still most
 probably involved in \emph{strong} three-dimensionality, but both \emph{weak} 
and \emph{strong} three-dimensionality only vanish in the limit 
$N_t'\rightarrow\infty$, and not at a critical value of $N_t'$. 
\item A cutoff scale exists in MHD turbulence, that separates quasi-two-dimensional scales from three-dimensional ones, as predicted by \cite{sm82}. Its
 existence could, however, only be confirmed experimentally for \emph{strong} 
three-dimensionality, even though it is expected to be found for \emph{weak} 
three-dimensionality too.
\end{itemize}
%

These results give a good idea of the conditions in which three-dimensionality 
is to be expected and how it arises. Since no direct measurements in the bulk were 
available, the question of how the appearance of three-dimensionality relates 
to that of the component of velocity along $\mathbf B$  was left aside in this 
particular paper. Its link to weak three-dimensionality was investigated  
elsewhere in MHD and non-MHD flows (\cite{davidson02_ejb,p12_epl, prcd13_epje}). Two complementary approaches are currently underway to obtain a precise diagnosis of the flow in the bulk: one based on numerical simulations of the experiment in its exact 
configuration, and the other using ultrasound velocimetry to measure velocity profiles directly. 
The question that now remains concerns how the different forms of 
three-dimensionality identified affect the flow dynamics, 
the finer properties of turbulence and in particular the direction of the 
energy cascade.\\

The authors gratefully acknowledge financial support from the Deutsche Forchungsgemeinschaft (grant PO1210/4-1). They would also like to thank an anonymous referee for the thoroughness and exemplary spirit of their report. 
\section{Additional remarks \label{sec:remarks}}
\begin{enumerate}
\item \emph{Intensity of Turbulent fluctuations \label{rem:fluc}}\\
Physically, the increase of $U'_b/U_b$ with $\lambda_i$ and $H\!a$
for a given value of $Re^0$ can be understood as follows: turbulent
structures become more and
more two dimensional when $H\!a$ is increased, and all the more so as they are
large (hence the positive exponent of $\lambda_i$). Energy transfer from the
main flow to them, and through them to smaller scales, is then progressively
impeded by friction in the Hartmann layers (which incurs a dissipation of the
order of $-H\!a \rho\nu U_b^{\prime 2}/l_\perp^2$, on a structure of size $l_\perp$),
rather than by Joule dissipation in the bulk of the flow (of the order of $-H\!a^2 \rho\nu U_b^{\prime 2}/l_\perp^2$). With less and less dissipation as the flow becomes closer to two dimensionality, more energy is retained by turbulent fluctuations. This effect can also be noticed in high-precision simulations of the flow in a duct past a cylindrical obstacle by \cite{kanaris13_pf}.

\item\emph{Validity of Electric Potential Velocimetry in strongly three-dimensional flows \label{rem:ept}}\\
A remark must be made here on the validity of the velocity measurements near
the top wall in such a strongly three-dimensional flow. Since the current
density in the core is comparable to that in the top Hartmann layer,
(\ref{eq:up_exp}) becomes inaccurate near the top wall. The quantity
$B^{-1}\overline{\langle|\nabla\phi_t|\rangle}$ may still be interpreted as a
velocity but represents an average over the upper layer of thickness
$h-l_z^{(N)}$, rather than the velocity in the vicinity of the top Hartmann layer.
\item\emph{$z$-linear variations of electric potential in the Shercliff layers \label{rem:shercl}}\\
A $z-$linear antisymmetric
component of $\phi_S(z)$, by contrast, corresponds to constant and therefore
non-divergent vertical current. Such a current can only be fed by
the Hartmann layers, which connect to the Shercliff layers at
the top and bottom edges of the vessel, not from the core.
\end{enumerate}

\bibliographystyle{jfm}
\bibliography{fullbiblio.bib}

\end{document}